\documentclass[10pt,letterpaper]{article}
\usepackage[top=0.85in,left=2.75in,footskip=0.75in]{geometry}
\usepackage[graphicx]{realboxes}

\usepackage{amsmath,amssymb}

\usepackage{changepage}

\usepackage[normalem]{ulem}
\useunder{\uline}{\ul}{}
\usepackage[utf8x]{inputenc}
\usepackage{subcaption}
\usepackage{mwe}
\usepackage{subcaption}
\usepackage{textcomp,marvosym}

\usepackage{cite}
\usepackage{float}
\usepackage{nameref,hyperref}

\usepackage[right]{lineno}

\usepackage{amsmath}
\usepackage{graphicx}
\usepackage{subcaption}
\usepackage{mwe}
\usepackage{microtype}
\DisableLigatures[f]{encoding = *, family = * }

\usepackage[table]{xcolor}

\usepackage{array}

\newcolumntype{+}{!{\vrule width 2pt}}

\newlength\savedwidth

\raggedright
\usepackage[aboveskip=1pt,labelfont=bf,labelsep=period,justification=raggedright,singlelinecheck=off]{caption}

\makeatletter
\renewcommand{\@biblabel}[1]{\quad#1.}
\makeatother

\usepackage{lastpage,fancyhdr,graphicx}
\usepackage{epstopdf}
\fancyheadoffset[L]{2.25in}
\fancyfootoffset[L]{2.25in}
\begin{document}
\vspace*{0.2in}

\begin{flushleft}
{\Large
\textbf\newline{
Characterizing Engagement Dynamics across Topics on Facebook} 

%

}


Gabriele Etta\textsuperscript{1},
Emanuele Sangiorgio\textsuperscript{2},
Niccolò Di Marco\textsuperscript{3},
Michele Avalle\textsuperscript{1},
Antonio Scala\textsuperscript{4},
Matteo Cinelli\textsuperscript{1},
Walter Quattrociocchi\textsuperscript{1}
\bigskip

\textbf{1} Center of Data Science and Complexity for Society, Department of Computer Science, Sapienza Università di Roma
\\
\textbf{2} Department of Social and Economic Sciences, Sapienza Università di Roma\\
\textbf{3} Department of Mathematics and Computer Science, University of Florence, Italy \\
\textbf{4} ISC-CNR UoS Sapienza, Rome,Italy \\
\bigskip

* quattrociocchi@di.uniroma1.it

\end{flushleft}

\section*{Abstract}

    Social media platforms heavily changed how users consume and digest information and, thus, how the popularity of topics evolves.
    In this paper, we explore the interplay between the virality of controversial topics and how they may trigger heated discussions and eventually increase users' polarization. 
    We perform a quantitative analysis on Facebook by collecting $\sim57M$ posts from $\sim2M$ pages and groups between 2018 and 2022, focusing on engaging topics involving scandals, tragedies, and social and political issues.
    Using logistic functions, we quantitatively assess the evolution of these topics finding similar patterns in their engagement dynamics.
    Finally, we show that initial burstiness may predict the rise of users' future adverse reactions regardless of the discussed topic.


\section*{Introduction}

The advent of social media platforms changed how users consume information online \cite{yasseri2022collective, lazaroiu2014role, ahmad2010twitter, notarmuzi2022universality}. 
The micro-blogging features on Twitter and Facebook, combined with a direct interaction between news producers and consumers, have remarkably affected how people get informed, shape their own opinions, and debate with other peers online \cite{brown2007word, kahn2004new, mcgregor2019social}. 
Over the years, following the business model of social media platforms, news outlets and producers attempted to maximize the time spent by users on their contents \cite{jaakonmaki2017impact, di2016social}, giving birth to the concept of {\em attention economy} \cite{simon1971designing}. The term refers to the users' limited capability and time to process all information they interact with \cite{kies2018social, holt2013age, brooks2015does}.
The transition toward a news ecosystem shaped on social media platforms unveiled patterns in information consumption at multiple scales \cite{cinelli2020selective, del2016spreading}, which contributed to the emergence of the polarisation phenomenon, and the formation of like-minded groups called echo chambers \cite{flaxman2016filter, cinelli2021echo, cookson2022echo}. Within echo chambers, characterized by homophily in the interaction network and bias in information diffusion towards like-minded peers, selective exposure \cite{klapper1960effects} is a significant driver for news consumption \cite{cinelli2021echo}. The combination of echo chambers and selective exposure makes users more likely to ignore dissenting information \cite{zollo2017debunking}, choosing to interact with narratives adhering to their point of view \cite{del2016spreading,
bessi2014economy}. 

Several studies explored the existence of these mechanisms in many topics concerning political elections, public health, climate change, and trustworthiness of the news sources  \cite{del2016spreading,bessi2014economy,mocanu2015collective,cinelli2020covid,etta2022covid,Falkenberg22,candia2019universal,briand2021infodemics,bovet2019influence,valensise2021lack}. 
Findings indicate neither the topic nor the quality of information explains the users' opinion-formation process. Instead, 
several studies observed how the virality of discussions can increase the likelihood of inducing polarization, hate speech, and toxic behaviors \cite{chang2021bat, cinelli2021dynamics, persily_tucker_2020}, highlighting how recommendation algorithms may have a role in shaping the news diet of users. 

Therefore, it is necessary to provide a better understanding of how user interest evolves in online debates.
To achieve this goal, we provide a quantitative assessment of the dynamics underlying user interest in news articles about different topics. In this paper, we 
analyze the engagement patterns produced by $\sim57M$ posts on Facebook related to $\sim300$ topics, involving a total of $\sim2M$ posting pages and groups over a period that ranges from $2018$ to $2022$. 
We first provide a quantitative assessment of topics' resonance through time, extracting insightful parameters from their engagement evolution. Then, we exploit the obtained parameters by assessing relationships with the sentiment expressed by users through their positive and negative reactions. Our results show that topics are generally characterized by an interest that constantly increases since the appearance of the first post. 
We find that topics' interactions grow with permanent intensity, even for prolonged periods, indicating how interest is a cumulative process that takes time.
We statistically validate this result by comparing parameters across topic categories, discovering no differences in the evolution of the engagement. Indeed, regardless of their category, topics keep users engaged steadily over time, and their lifetime progression seems thus unrelated to its thematic field. 
Finally, we find that topics with sudden virality tend to trigger more controversial and heterogeneous interactions. In turn, topics with a steady evolution exhibit more positive and homogeneous reaction types. 
This difference in the sentiment of reactions, and the protracted duration of topics' lifetime, are both upshots consistent with the emergence of selective exposure as a driver of news consumption.

\section*{Materials and Methods}
\label{sec:methods}
This section describes the data collection process, the topic extraction process, the models and the metrics employed in assessing collective attention.

\subsection*{Overview of the data collection process}
\begin{figure}[ht!]
    \centering
    \includegraphics[scale = 0.23]{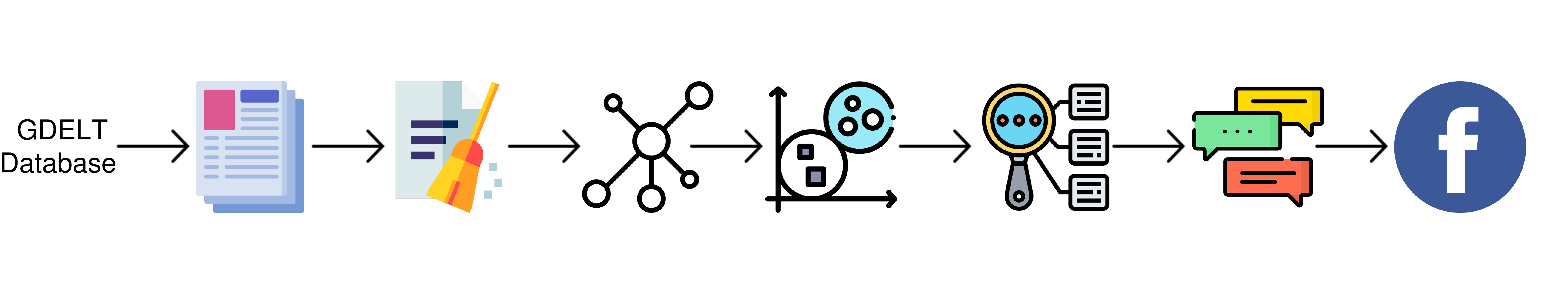}
    \caption{Summary of the analysis workflow followed in the current study. News articles are collected from the GDELT Database, and their corpus is extracted, cleaned and analyzed to retrieve the most representing terms. The co-occurrence network built upon these terms serves as an input for the Louvain community detection algorithm to identify keyword clusters. Independent labellers then analyze these clusters to identify the subset of words that represent the topic under consideration, which are then used on Crowdtangle to retrieve the Facebook posts relating to those events.}
    \label{fig:analysis_pipeline}
\end{figure}

The data collection process comprises several parts, as described in Fig.~\ref{fig:analysis_pipeline}. We start by creating a sample of news articles from the GDELT event database \cite{gdelt_event_database}, and then we process the articles' text to obtain a set of representing terms. Consequently, we apply the Louvain community detection algorithm \cite{blondel2008fast} on the co-occurrence term network to identify the topics of interest. The terms representing these topics will serve as input for the collection of posts from Facebook.

\subsubsection*{News Extraction from GDELT}
The GDELT (Global Database of Events, Language, and Tone) Project \cite{gdelt_project}, powered by Google Jigsaw, is a database of global human society which monitors the world's broadcast, print, and web news from nearly every corner of every country in more than 100 languages. It identifies the people, locations, organisations, themes, sources, emotions, counts, quotes, images and events driving our global society every second of every day \cite{leetaru2013gdelt}. We gathered news articles from the GDELT 2.0 Event Database \cite{gdelt_event_database}, which can store new world's breaking events every 15 minutes and translates the corresponding news articles in 65 languages, representing 98.4\% of its daily non-English monitoring volume \cite{gdelt_translingual}. The analysis covers a period between $1/1/2018$ and $13/5/2022$, collecting $50$ news articles each week for a total of $\sim79K$. 

\subsubsection*{Extracting representative keywords from news articles}
To clean and extract the most representative keywords of each news article, we employed the \textit{newspaper3k} Python package \cite{newspaper3k}. We initially extracted words from the body of the article, excluding stopwords and numbers. Then, we computed the word frequency $f(w,i)$ for each word $w$ in article $i$. Finally, we sorted words in descending order according to their frequency, keeping the top 10 most frequent words.

\subsubsection*{Topic Extraction from News Article's Keywords}
\label{sec:topics}
The list of terms with the corresponding news articles can be formalised as a bipartite graph $G = (T, A, E)$ whose partitions $T$ and $A$ represent the set of terms $t \in T$ and the articles $a \in A$ respectively, for which an edge $(t, a) \in E$ exists if a term $t$ is present in an article $a$. By projecting graph G on its terms $T$ we obtain an undirected graph $P$ made up of nodes $t \in T$, which are connected if they share at least one news article.\\
We perform community detection on the nodes of $P$ by employing the Louvain algorithm \cite{blondel2008fast}. As a result, we obtain a set of clusters $C$, where each cluster $c \in C$ contains a list of keywords that are assumed to be semantically related to a topic. We then asked a pool of three human labellers to select, for each community, from two to three terms they considered the most representative to identify a topic unambiguously.

\subsubsection*{Data collection of Facebook posts}
\label{sec:fb_data}
The news articles obtained from the GDELT Event Database do not contain information helpful in estimating the attention they generate online. 
To include the dimension of user engagement, we employ each topic's set of representative terms to collect Facebook data over a period that goes from $01/01/2018$ to $05/05/2022$. The data was obtained using CrowdTangle \cite{crowdtangle}, a Facebook-owned tool that tracks interactions on public content from Facebook pages, groups, and verified profiles. CrowdTangle does not include paid ads unless those ads began as organic, non-paid posts that were subsequently ``boosted'' using Facebook's advertising tools. CrowdTangle also does not store data regarding the activity of private accounts or posts made visible only to specific groups of followers. 

The collection process produced a total of $\sim57M$ posts from  $\sim2M$ unique pages and groups, generating $\sim 8B$ interactions. The result of the data collection process is described in Table \ref{tab:data_breakdown}.

\begin{table}[H]
\resizebox{1\textwidth}{!}{%
\begin{tabular}{|l|l|l|l|l|l|}
\hline
\textbf{\begin{tabular}[c]{@{}l@{}}Total News Articles\\from GDELT\end{tabular}} &\textbf{\begin{tabular}[c]{@{}l@{}}Total Posts \\from Facebook\end{tabular}} & \textbf{\begin{tabular}[c]{@{}l@{}}Total\\ Interactions\end{tabular}} & \textbf{\begin{tabular}[c]{@{}l@{}}Total Groups \\ and Pages\end{tabular}} & \textbf{\begin{tabular}[c]{@{}l@{}}Number of Topics\\ Collected\end{tabular}} & \textbf{Period}      \\ \hline
79 650 & 57 031 026
                                                   & 8 015 177 602
                                                      &      2 224 430                                                        & 296                                                                           & 1/1/2018 - 13/5/2022 \\ \hline
\end{tabular}
}

\caption{Data Breakdown of the study, including the total amount of news articles and posts collected from GDELT and Facebook respectively, together with the number of topics and the analysis period..}
\label{tab:data_breakdown}
\end{table}

\subsubsection*{Topic Categorization}
\label{sec:topic_categorization}
To provide a correspondence between topics and their area of interest, we performed a categorization activity under the following labels: Art-Culture-Sport (ACS), Economy, Environment, Health,
Human Rights, Labor, Politics, Religion, Social and Tech-Science. Three human labellers carried out the activity to connect topics and categories, choosing as the representative only those categories selected by at least two of the three labellers. 

\subsection*{Metrics}
We begin by describing a measure for fitting the cumulative engagement evolution. Then, based on the previous step, we outline an index to evaluate the sharpness of the topic's diffusion. Finally, we introduce a sentiment score to assess the topic's controversy by using Facebook's reactions.

\subsubsection*{Fitting cumulative engagement evolution}
\label{sec:sigmoid}
The diffusion of new ideas has been widely studied in the past \cite{de1903laws, rogers1976product, grubler1990rise, perez2003technological, robinson2012changeology, kanjanatarakul2012comparison}, indicating how the logistic function can effectively model the diffusion of innovations. Therefore, to model the evolution of the engagement received by posts, we fit the cumulative distribution of the overall engagement ( i.e., the number of likes, shares and comments) over time employing a function $f_{\alpha, \beta} (t)$, with $\alpha, \beta \in \mathbb{R}$, defined as

\begin{equation}
\label{eq:sigmoid}
  f_{\alpha, \beta} (t) = \frac{1}{1+e^{-\alpha \left(t-\beta\right)}}.
\end{equation}

From a mathematical point of view, Eq. \ref{eq:sigmoid} defines a general sigmoid function that depends on the parameters $\alpha$ and $\beta$. The $\alpha$ parameter represents the slope of the function, describing the steepness of the engagement evolution. On the other hand, $\beta$ is the point at which the function reaches the value $0.5$ and quantifies the time required for a topic to reach half its total interactions.

\newpage
    \begin{figure*}
        \centering
        \begin{subfigure}[b]{0.47\textwidth}
            \centering
            \includegraphics[width=\textwidth]{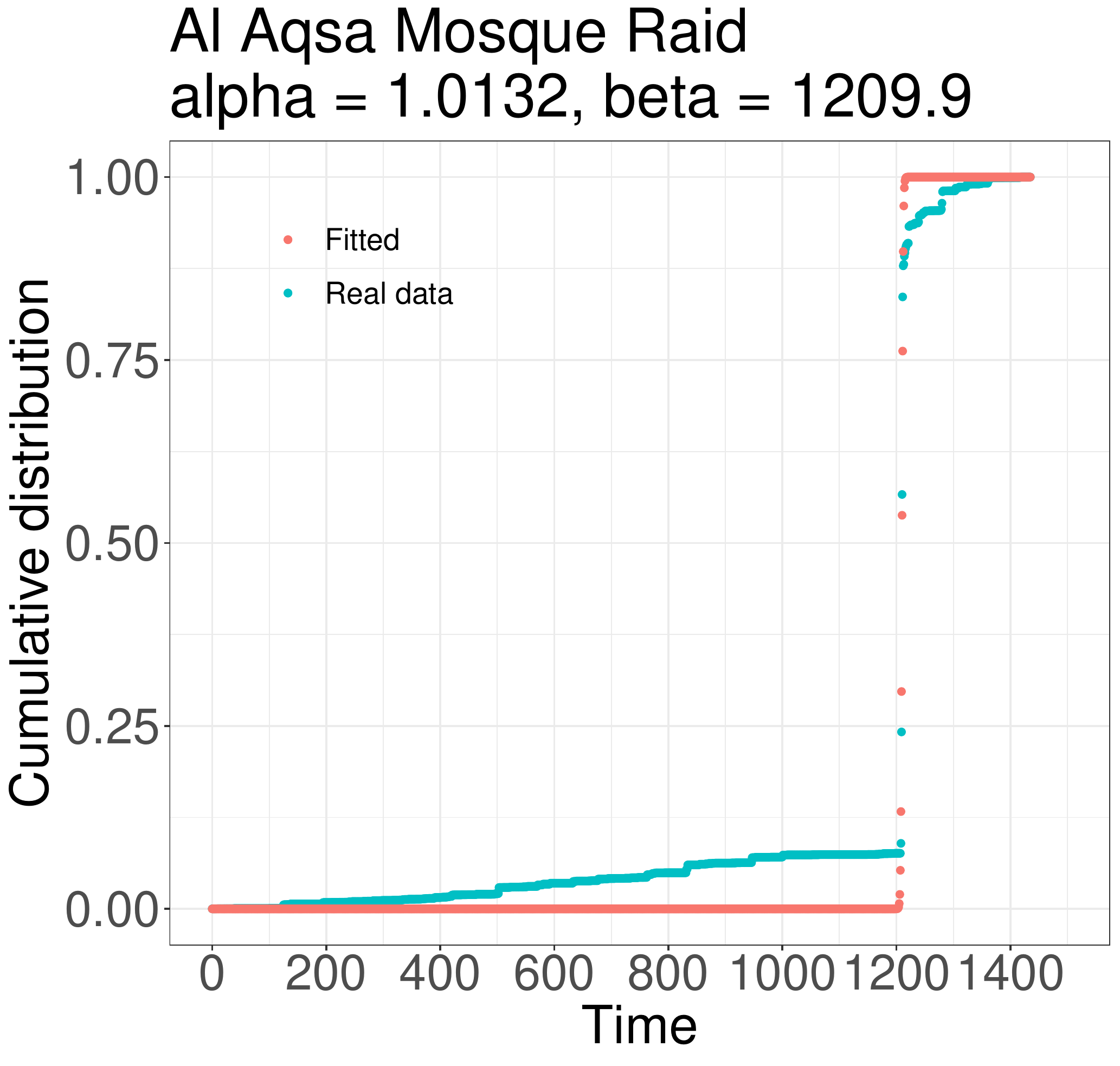}
            \caption{}
            \label{fig:cumulative_mosque}
        \end{subfigure}
        \hfill
        \begin{subfigure}[b]{0.47\textwidth}  
            \centering 
            \includegraphics[width=\textwidth]{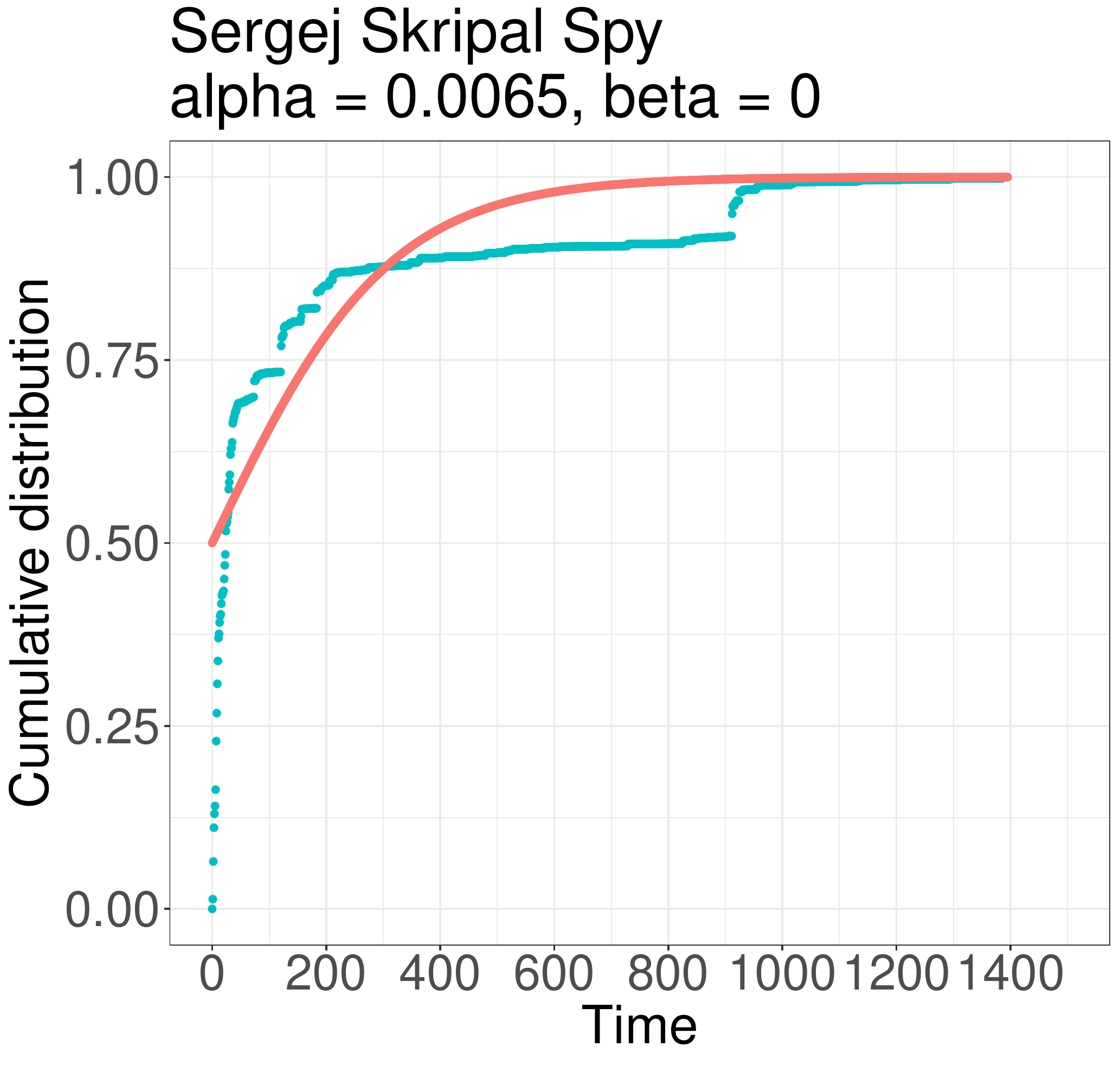}
            \caption{}
            \label{fig:cumulative_skipral}
        \end{subfigure}
        \begin{subfigure}[b]{0.47\textwidth}   
            \centering 
            \includegraphics[width=\textwidth]{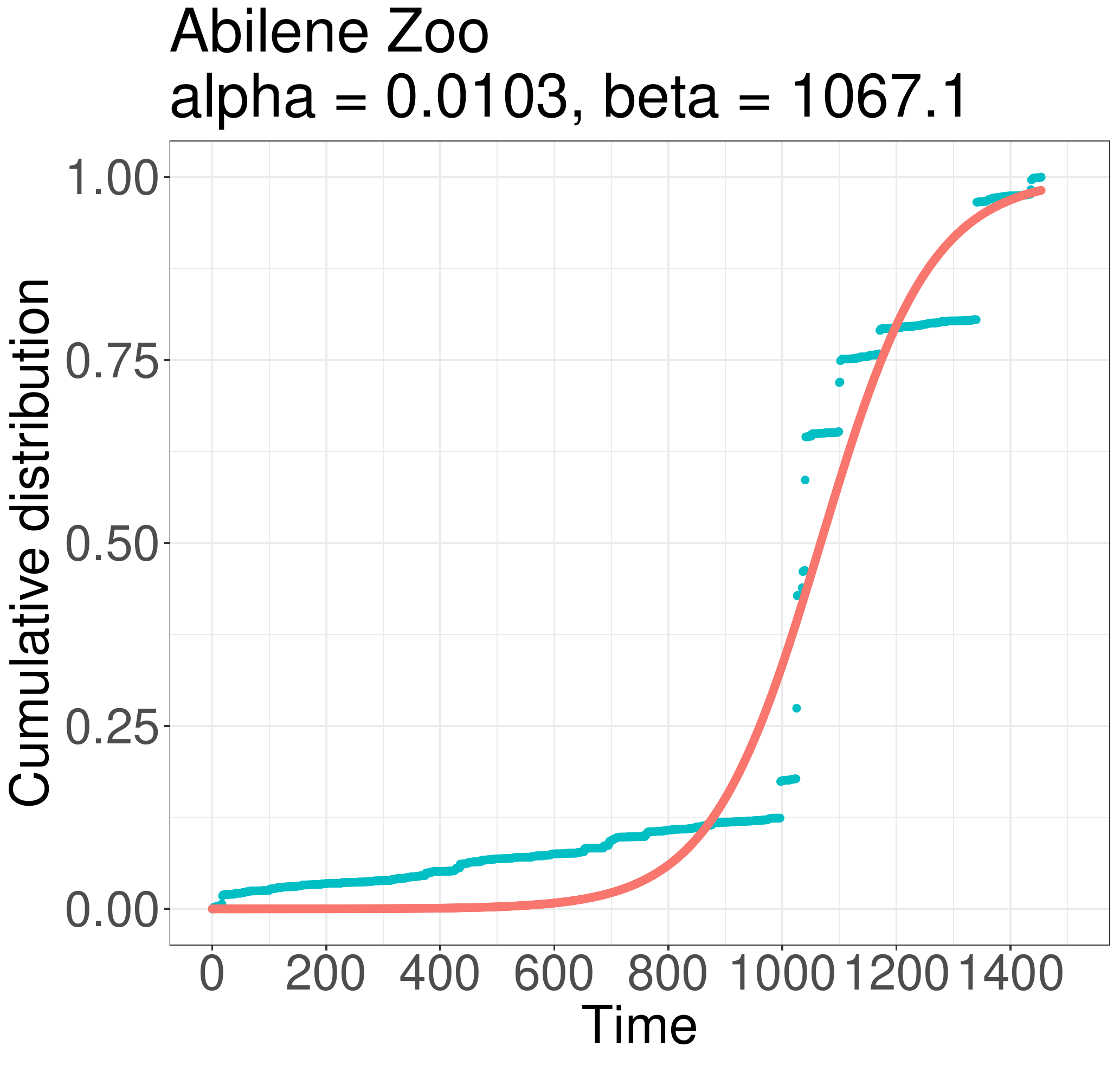}
            \caption{ }
            \label{fig:cumulative_abilene}
        \end{subfigure}
        \hfill
        \begin{subfigure}[b]{0.47\textwidth}   
            \centering 
            \includegraphics[width=\textwidth]{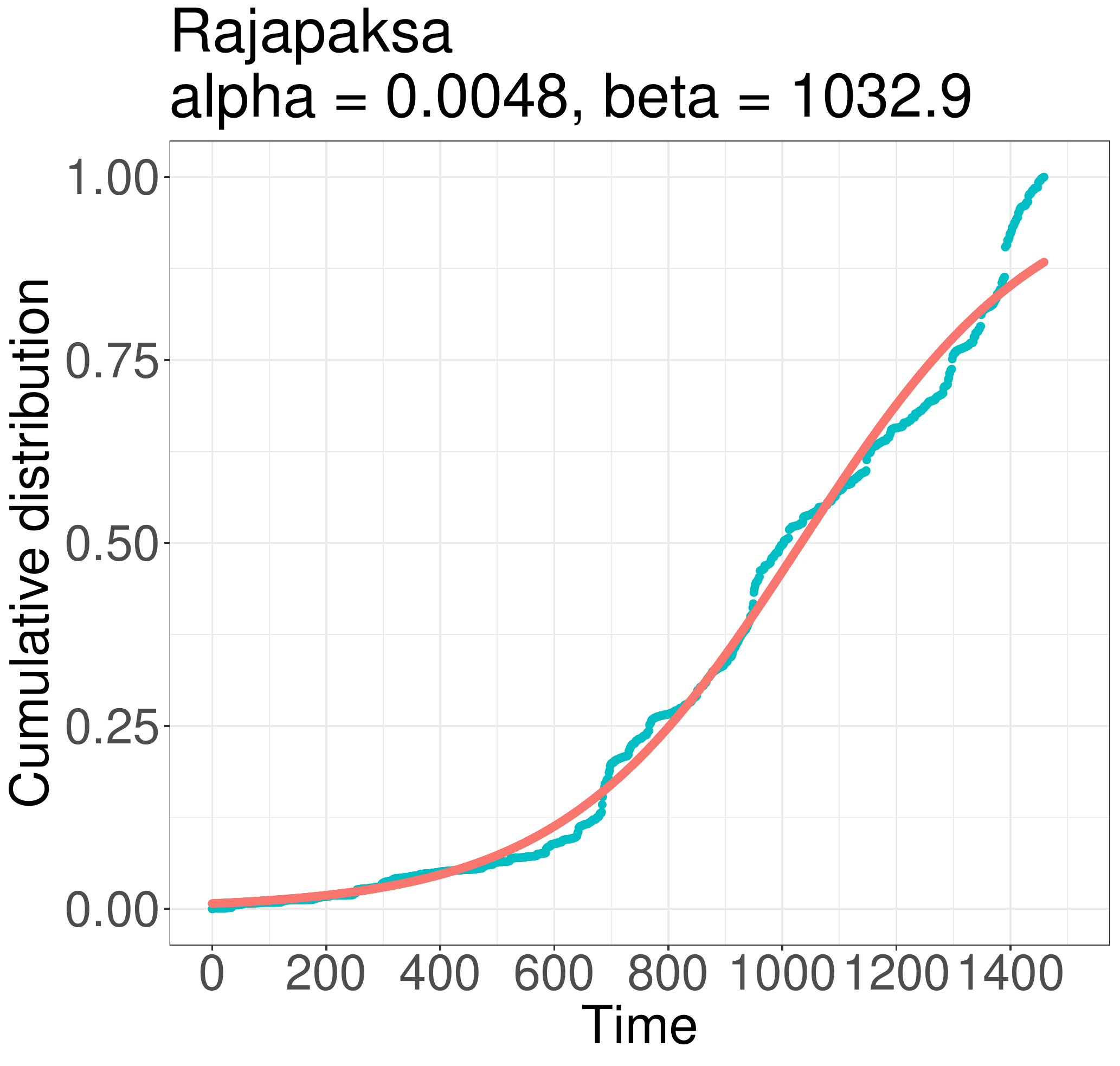}
            \caption{ }
            \label{fig:cumulative_rajapaksa}
        \end{subfigure}
        \caption{Representation of a sample of four topics employing their normalized cumulative evolution of engagements and fittings. The incidence of the $\alpha$ parameter can be observed in the sharpness of the fitting curves. The $\beta$ parameter instead regulates the shift of the function through the $x$ axis: the higher its value, the higher the delay from $t_0$ where the sigmoid produces its increment.}
        \label{fig:fitting_examples}
    \end{figure*} 

To provide a representation of the impact that $\alpha$ and $\beta$ can have in topic engagement evolution, Fig. \ref{fig:fitting_examples} displays four topics with peculiar configurations. Fig. \ref{fig:cumulative_mosque} shows a sigmoid in which the high values of $\alpha$ and $\beta$ produce a sharp increment relatively far from $t_0$. Such behaviour corresponds to those topics that require some time before gaining resonance with the public. Fig. \ref{fig:cumulative_skipral} instead provides a fit where the sigmoid produces low values for $\alpha$ and $\beta$, resulting in a smoother increment in the proximity of $t_0$ than the one described in Fig.\ref{fig:cumulative_mosque}. Finally, Fig.  \ref{fig:cumulative_abilene} and \ref{fig:cumulative_rajapaksa} provide an example of how two curves that share similar values of $\beta$ parameters can have a different evolution of their increase by slightly modifying the values for $\alpha$ parameter.

\subsubsection*{Speed Index}
\label{sec:speed_index}
To model the evolution of a topic by taking into account the joint contribution of $\alpha$ and $\beta$ parameters, we define a measure called the Speed Index
$SI(f_{\alpha,\beta})$ as

\begin{equation}
\label{eq:speed_index}
SI(f_{\alpha,\beta}) = \frac{\int_{0}^T f_{\alpha,\beta}(t) dt}{T},
\end{equation}

where $T$ represents the time of the last observed value for $f_{\alpha,\beta} (t)$. Note that $SI$ is the mean integral value of $f_{\alpha,\beta}$, i.e. the normalised area under the curve of $f_{\alpha,\beta}$ (therefore $SI(f_{\alpha,\beta}) \in [0,1]$). The assumption in the definition of this function relies on the fact that high-speed values are obtained by sigmoids that reach the plateau in a short time, as the behaviour represented in Fig. \ref{fig:cumulative_skipral}.

\subsubsection*{Love-Hate Score}
\label{sec:lhi}
To quantify the level of sentiment that a Facebook post produces, we define a measure of controversy called Love-Hate Score $LH (i) \in \left[-1,1 \right]$ as

\begin{equation}
\label{eq:lhi}
LH (i) = \frac{l_i-h_i}{l_i+h_i},
\end{equation}

where $h_i$ and $l_i$ are respectively the total number of \textit{Angry} and \textit{Love} reactions collected by a post $i$. A value of $LH$ equal to $-1$ indicates that the post received only \textit{Angry} reactions from the users, while a value equal to $1$ indicates that the post received only $Love$ reactions.

\section*{Results and Discussion}

\subsection*{Quantifying topic resonance}
\label{sec:resonance_distribution}
We first provide a quantitative assessment of the topics' resonance on social media. To do so, we perform a Non-linear Least Squares (NLS) regression by fitting the sigmoid function $f_{\alpha,\beta} (t)$ to the cumulative evolution of the engagement for each topic.

\begin{figure}[H]
    \centering
    \includegraphics[scale = 0.35]{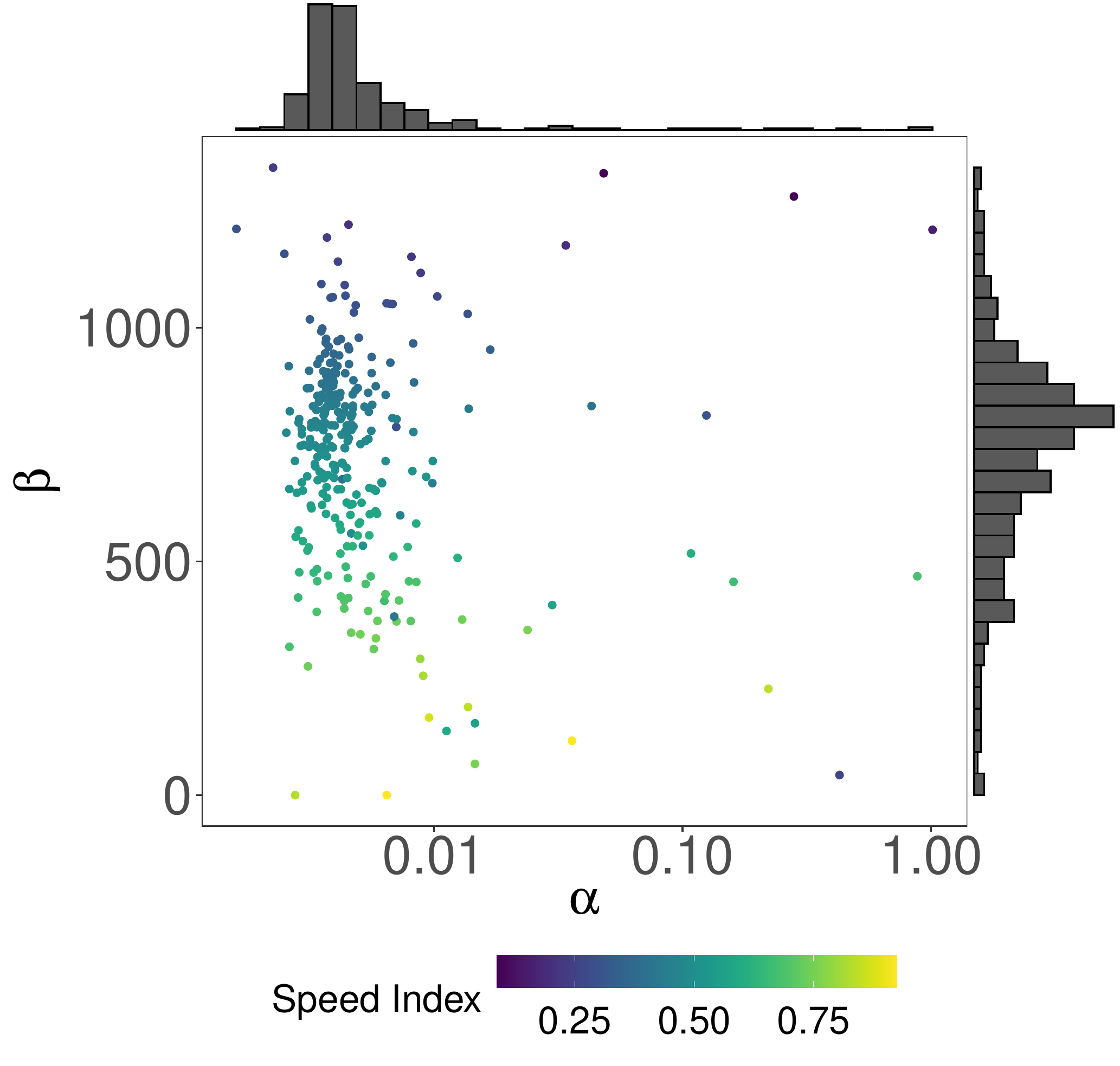}
    \caption{Joint distribution of $\alpha$ and $\beta$ parameters obtained from the NLS regression for each topic. We observe that topics are generally characterized by values of $\alpha$ and $\beta$, which explains how user interest in a topic does not increase all of a sudden but is the result of a process that evolves over time.}
    \label{fig:joint_parameter_distribution}
\end{figure}

The distribution of the $\alpha$ parameter provided in Fig. \ref{fig:joint_parameter_distribution} describes how the majority of topics have a value of $\alpha$ belonging to the $\left[0, 0.0047 \right]$ interval. This result demonstrates how user interest in a topic does not suddenly increase but results from a long-term process. The distribution of the $\beta$ parameter, instead, describes a prevalence of topics in the $\left[600,1000 \right]$ interval, identifying the tendency of topics to become a matter of interest with some delay w.r.t the first post covering them.

\subsection*{Evaluating the relationship between topic resonance and controversy}
\label{sec:resonance_sentiment_relationship}

\begin{figure}[H]
    \centering
    \includegraphics[scale = 0.25]{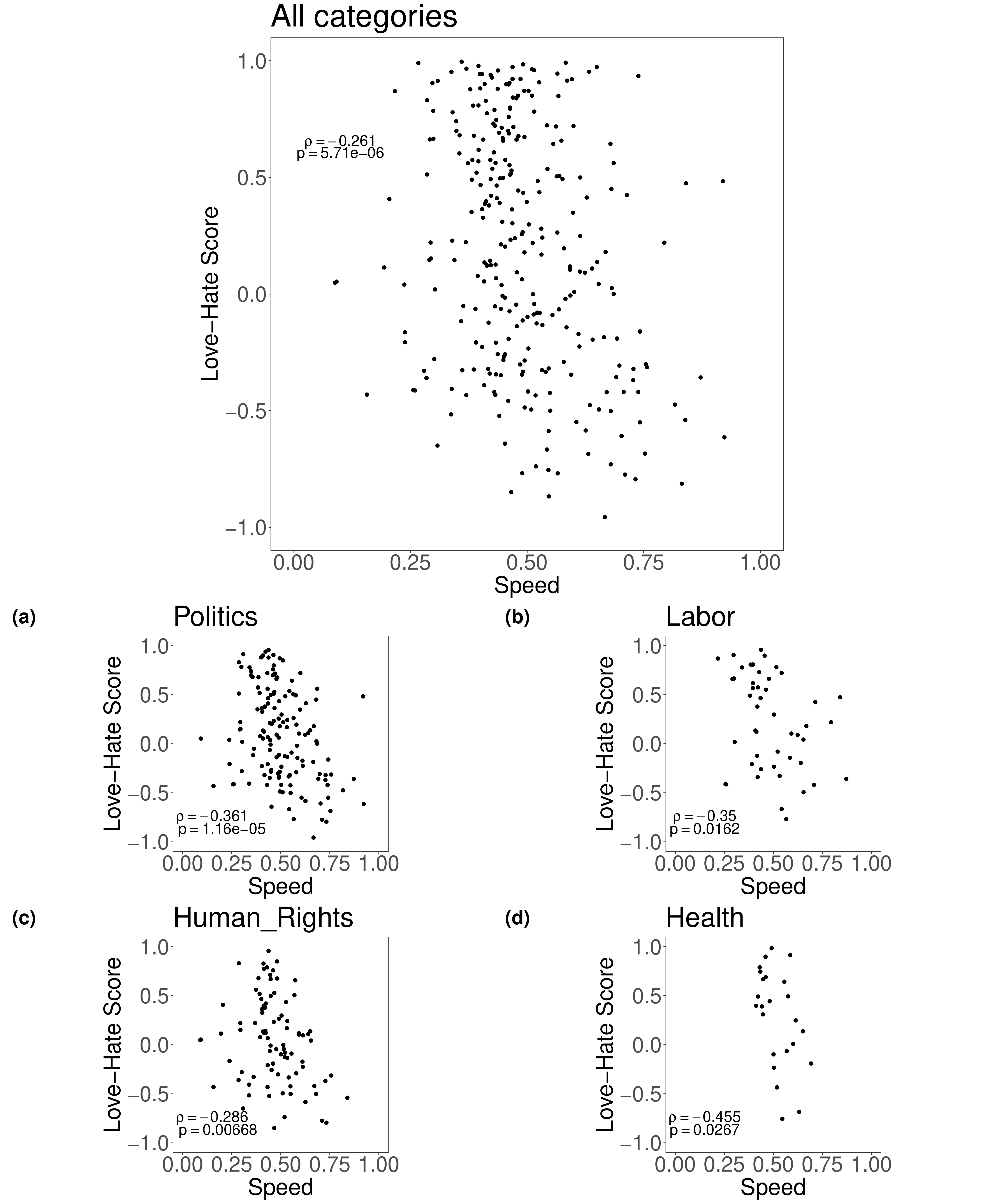}
    \caption{Upper panel: correlation between $SI$ and $LH$ score for each of the topics identified. Lower panel: correlation between $SI$ and $LH$ score for the top $4$ most frequent topics. Overall, we observe how users react negatively as topics become sharply viral. 
    }
    \label{fig:LH_speed}
\end{figure}

\begin{figure}[!ht]
    \centering
    \includegraphics[scale = 0.35]{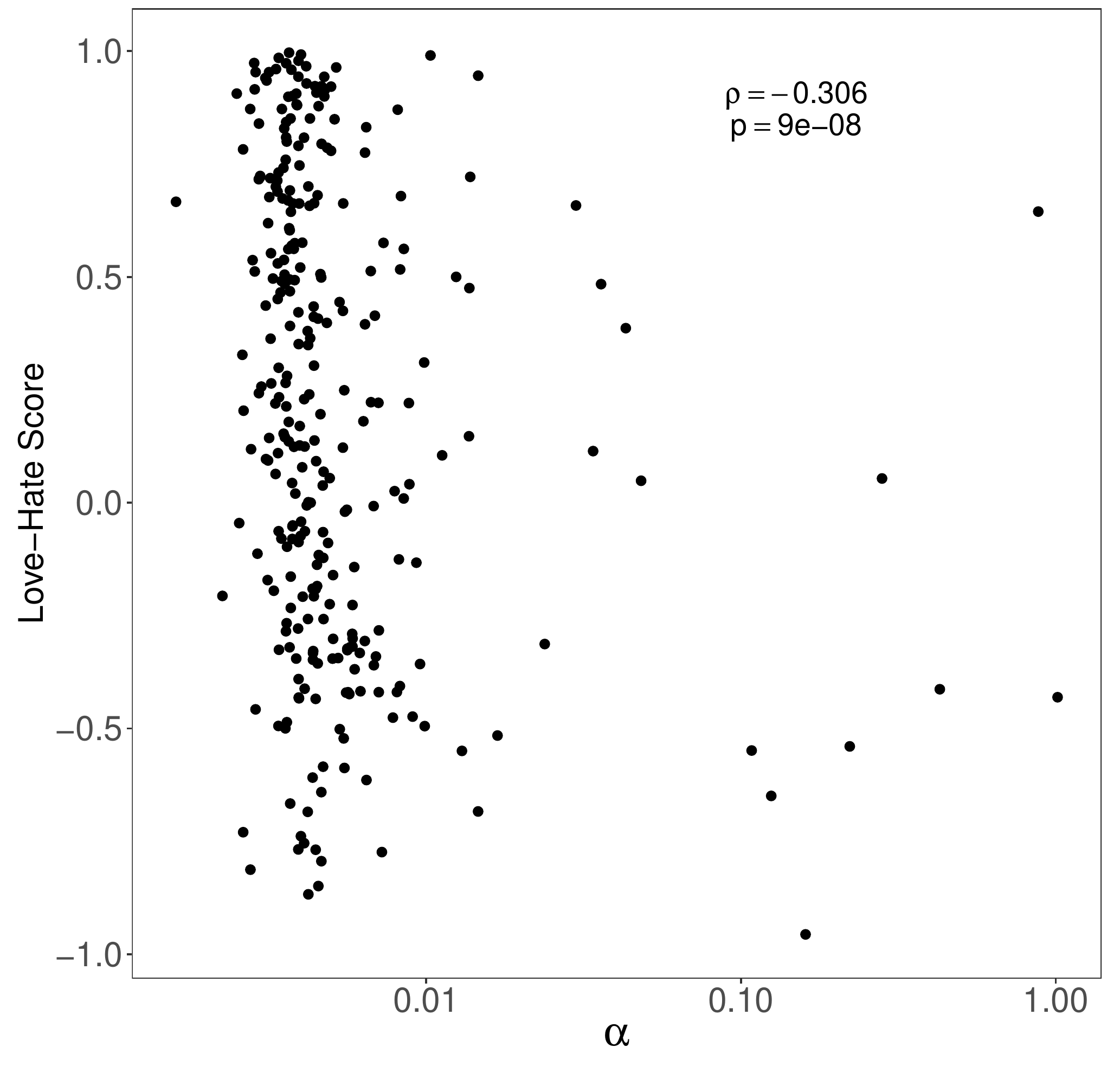}
    \caption{Correlation between $\alpha$ and $LH$ score for each of the topics identified}
    \label{fig:cor_alpha_LH}
\end{figure}

To quantify the interplay between user interest in a topic and the controversy it produces, we compute the Spearman correlation between the Speed Index and the Love-Hate (LH) Score for each topic. Results from the upper panel of Fig. \ref{fig:LH_speed} show a general negative tendency of users to react with adverse sentiment when a topic gains engagement faster ($\rho = -0.26$), leaving positive reactions to those topics that require time to gain resonance. Results described in the lower panel of Fig. \ref{fig:LH_speed} provide further characterization of the interplay between the Speed Index and the Love-Hate score after classifying the topics according to the four most frequent categories analyzed, i.e., Politics, Labor, Human Rights and Health. We observe how the Politics and Health categories have the lowest correlation scores ($\rho = -0.36$ and $\rho = -0.45$), providing further evidence of their intrinsic polarizing attitude (see Table \ref{tab:corr_speed_LH} for the complete list of correlation coefficients). Furthermore, the correlation between $\alpha$ and LH Score produces similar results as with the Speed Index (see Fig. \ref{fig:cor_alpha_LH} in SI for more details). 

\subsection*{Assessing the differences of engagement behaviors across topic categories}
To conclude our analysis, we investigate the differences in the evolution of engagement across topic categories. In particular, for each parameter distribution ($\alpha$, $\beta$ and $SI$), we apply a two-tailed Mann–Whitney U test \cite{mann1947test} to each pair of parameters. Table ~\ref{tab:test_results} provides the percentages of the significant p-values for the four parameters. Due to the necessity to perform multiple tests, we apply a Bonferroni correction to our standard significance level of $0.05$, leading to reject the null hypothesis if the p-value $p < 0.001$. 
Our results show that the resulting p-values from the tests do not lead to rejecting the null hypothesis. Such a result corroborates the hypothesis that, on average, users are characterized by homogeneous engagement patterns that are not influenced by the consumed topic.
We further extend the statistical assessment by performing the same
test between Love-Hate Score distributions of the different categories. 

\begin{table}[H]
\centering
\begin{tabular}{|l|l|l|l|l|}
\hline
                            & \textbf{$\alpha$} & \textbf{$\beta$} & \textbf{Speed Index} & \textbf{Love-Hate} \\ \hline
\textbf{\textless{}0.001}    & 2.22\%           & 0\%          & 0\%                 & 20\%               \\ \hline
\textbf{\textgreater{}0.001} & 97.78\%           & 100\%          & 100\%                 & 80\%               \\ \hline
\end{tabular}
\caption{Percentage of p-values resulting from the two-sided Mann–Whitney U test between each category employing their $\alpha$, $\beta$, Speed Index and Love-Hate Score.}
\label{tab:test_results}
\end{table}

Conversely to engagement evolution results, the topic's category explains differences in the sentiment of reactions in 20\% of cases. 
Such findings reveal that some categories are composed of significantly more negative and controversial topics, indicating how elicited reactions vary according to specific subjects. Understanding that some of them are more prone to induce negative feedback from users could be a proxy to introduce their related topics in the online debate.


\section*{Conclusions}
In this work, we perform a quantitative analysis of user interest on a total of $\sim57M$ Facebook posts referring to $\sim300$ different topics ranging from $2018$ to $2022$. We initially quantify the distribution of topics' resonance throughout the analysis. Then, we evaluate the relationship between engagement and controversy. Ultimately, we assess the differences in engagement across different categories of topics.
Our findings show that, on average, user interest in topics does not increase exponentially right after their appearance but, instead, it grows steadily until it reaches a saturation point.
From a sentiment perspective, topics that gained resonance right after their initial appearance are more likely to collect negative/controversial reactions, whilst topics which are more steady in their growth tend to attract positive user interactions. This result provides evidence about how recommendation algorithms should introduce topics adequately since sudden rises in topic resonance tend to reinforce polarization mechanisms.
Finally, we find no statistical difference between user interest across different categories of topics, providing evidence that, on a relatively large time window, the evolution of engagement with posts is primarily unrelated to their subject. On the contrary, we observe differences in the sentiment generated by the different topics, providing evidence of how polarisation drives people to perceive the piece of content they consume online in different ways, according to their framing and system of beliefs.

User interest and engagement evolution in the online debate are both aspects of human behaviour on social media whose underlying dynamics still need to be discovered from an individual point of view. Our findings provide an aggregate perspective of the interplay between major emerging behavioral dynamics and topics' lifetime progression, deepening the relationship between diffusion patterns and users' reactions.
Understanding that topics with an early burst in virality are associated with primarily adverse reactions from users sheds light on their tendency to react instinctively to new content. This approach enables the identification of highly polarizing topics, since their initial stage of diffusion, by observing the heterogeneity of users' reactions.
The following study presents some limitations. In data collection, CrowdTangle provides only posts from public Facebook pages with more than 25K Page Likes or Followers, public Facebook groups with at least 95K members, all US-based public groups with at least 2K members, and all verified profiles. These restrictions affected our datasets' sample and our findings' generality. Moreover, we did not have access to removed posts, groups, and pages, which could have been a meaningful proxy to characterize the attention dynamics of retracted content. Finally, since Crowdtangle does not provide information about users interacting with posts, we cannot assess their engagement from an individual perspective and model the possible relationship between users and topics employing a network approach.

Future works may extend the application of the proposed methodology on additional social media platforms to assess the role of the algorithms in the attention economy. Researchers may also take advantage of the extensions to further platforms by assessing the attention dynamics of users concerning specific topics.
\nolinenumbers

\section{Acknowledgements}
We acknowledge the 100683EPID Project “Global Health Security Academic Research Coalition” SCH-00001-3391.

\newpage
\section*{Supporting Information}
\label{sec:si}

\subsection*{List of topics employed}
\label{sec:list_topics}
\begin{table}[H]
\resizebox{1\textwidth}{!}{%
\begin{tabular}{|l|l|l|l|}
\hline
\textbf{Topic   Keywords}                          & \textbf{First Post Date} & \textbf{Last Post Date} & \textbf{Categories}                                            \\ \hline
Amyotrophic\_lateral\_sclerosis                    & 2018-01-02               & 2021-12-31              & Social, Health                                                 \\ \hline
DeleteUber                                         & 2018-01-01               & 2021-12-19              & Labor, Social                                                  \\ \hline
Roy\_Moore\_sexual\_misconduct                     & 2018-01-02               & 2021-12-13              & Human\_Rights, Politics, Social                                \\ \hline
abilene\_zoo                                       & 2018-01-08               & 2022-01-01              & Art\_Culture\_Sport, Environment                               \\ \hline
abu\_sayyaf                                        & 2018-01-02               & 2021-12-31              & Human\_Rights, Politics, Religion                              \\ \hline
action\_news\_jax                                  & 2018-01-01               & 2021-12-31              & Art\_Culture\_Sport                                            \\ \hline
afghan\_refugees                                   & 2018-01-01               & 2021-12-31              & Human\_Rights                                                  \\ \hline
afghanistan\_pakistani\_militant                   & 2018-01-02               & 2021-12-31              & Human\_Rights, Politics, Religion                              \\ \hline
afghanistan\_war                                   & 2018-01-02               & 2022-01-01              & Human\_Rights, Politics, Religion                              \\ \hline
afp\_paedophile\_ring                              & 2018-08-15               & 2021-11-02              & Human\_Rights                                                  \\ \hline
agent\_skripal\_spy                                & 2018-03-05               & 2021-12-29              & Politics                                                       \\ \hline
aids\_hiv                                          & 2018-01-02               & 2021-12-31              & Social, Health                                                 \\ \hline
al\_aqsa\_jerusalem\_raid                          & 2018-01-15               & 2021-12-20              & Human\_Rights, Politics, Religion, Social                      \\ \hline
alaska\_pipeline                                   & 2018-01-02               & 2021-12-31              & Economy, Environment                                           \\ \hline
alex\_jones                                        & 2018-01-02               & 2021-12-31              & Art\_Culture\_Sport, Politics, Social                          \\ \hline
alshabab\_mogadishu\_somalia                       & 2018-01-24               & 2021-12-31              & Human\_Rights, Politics, Religion                              \\ \hline
aluminium\_steel\_tariffs                          & 2018-01-11               & 2021-12-31              & Economy, Labor, Politics                                       \\ \hline
andhra\_pradesh\_uttarandhra                       & 2018-02-05               & 2021-12-03              & Economy, Politics, Social                                      \\ \hline
animal\_conservation                               & 2018-01-01               & 2021-12-31              & Environment                                                    \\ \hline
animal\_cruelty                                    & 2018-01-02               & 2021-12-31              & Environment, Health                                            \\ \hline
animal\_sanctuary\_tiverton                        & 2018-03-19               & 2021-12-18              & Environment, Labor                                             \\ \hline
antarctic\_ice\_melting                            & 2018-01-02               & 2021-12-31              & Environment, Social                                            \\ \hline
antisemitic\_jewish\_orthodox                      & 2018-02-02               & 2021-12-26              & Human\_Rights, Religion, Social                                \\ \hline
apc\_pdp\_sheriff                                  & 2018-01-03               & 2021-12-30              & Politics                                                       \\ \hline
armenia\_azerbaijan\_border                        & 2018-01-09               & 2021-12-31              & Politics, Social                                               \\ \hline
arvind\_kejriwal                                   & 2018-01-02               & 2020-12-31              & Human\_Rights, Politics, Social                                \\ \hline
ashland\_fundraising                               & 2018-01-03               & 2021-12-31              & Art\_Culture\_Sport, Economy, Social                           \\ \hline
asian\_hate                                        & 2018-01-01               & 2021-12-31              & Human\_Rights, Social                                          \\ \hline
aung\_san\_suu\_kyi                                & 2018-01-02               & 2021-12-31              & Human\_Rights, Politics, Social                                \\ \hline
australian\_refugees                               & 2018-01-02               & 2022-01-01              & Human\_Rights, Labor, Social                                   \\ \hline
baghdad\_shiites                                   & 2018-01-04               & 2021-11-29              & Religion, Social                                               \\ \hline
ballistic\_missile\_test                           & 2018-01-01               & 2021-12-31              & Environment, Politics, Social, Tech\_Sci                       \\ \hline
band\_debut\_album                                 & 2018-01-01               & 2021-12-31              & Art\_Culture\_Sport                                            \\ \hline
benghazi\_libya\_militias                          & 2018-01-04               & 2021-12-21              & Human\_Rights, Politics, Religion                              \\ \hline
bilateral\_cooperation                             & 2018-01-01               & 2021-12-31              & Economy, Politics                                              \\ \hline
birds\_invasive\_population\_species\_conservation & 2018-01-05               & 2021-12-31              & Environment                                                    \\ \hline
black\_racism                                      & 2018-01-01               & 2021-12-31              & Human\_Rights, Social                                          \\ \hline
blacklivesmatter                                   & 2019-05-21               & 2021-12-31              & Human\_Rights, Social                                          \\ \hline
blue\_whale\_challenge                             & 2018-01-01               & 2021-12-31              & Social, Health                                                 \\ \hline
boat\_sinks\_die                                   & 2018-01-01               & 2021-12-25              & Human\_Rights, Social                                          \\ \hline
boeing\_737\_max\_crash                            & 2018-01-01               & 2021-12-31              & Social                                                         \\ \hline
boko\_haram                                        & 2018-01-02               & 2022-01-01              & Human\_Rights, Politics, Religion                              \\ \hline
bollywood\_celebrities                             & 2018-01-01               & 2021-12-31              & Art\_Culture\_Sport                                            \\ \hline
bolsonaro\_brazil                                  & 2018-01-03               & 2021-12-31              & Human\_Rights, Politics, Social                                \\ \hline
bomber\_commits\_suicide                           & 2018-01-24               & 2021-12-14              & Social                                                         \\ \hline
boris\_hunt\_tory\_debate                          & 2018-01-31               & 2021-09-04              & Politics, Social                                               \\ \hline
boris\_johnson                                     & 2018-01-01               & 2021-12-31              & Politics                                                       \\ \hline
bowe\_bergdahl                                     & 2018-01-03               & 2021-12-26              & Human\_Rights                                                  \\ \hline
brain\_cells\_tumour                               & 2018-01-03               & 2021-12-31              & Social, Tech\_Sci, Health                                      \\ \hline
breast\_cancer                                     & 2018-01-01               & 2021-12-31              & Social, Health                                                 \\ \hline
britain\_bridge\_collapse                          & 2018-01-03               & 2021-12-23              & Art\_Culture\_Sport, Environment                               \\ \hline
bsf\_jammu\_kashmir                                & 2018-01-02               & 2021-12-31              & Politics                                                       \\ \hline
buckingham\_palace                                 & 2018-01-02               & 2021-12-31              & Art\_Culture\_Sport, Politics, Social                          \\ \hline
burqa                                              & 2018-01-01               & 2021-12-31              & Human\_Rights, Religion, Social                                \\ \hline
bus\_accident                                      & 2018-01-01               & 2021-12-31              & Labor, Social                                                  \\ \hline
\end{tabular}

}

\end{table}

\begin{table}[H]
\resizebox{1\textwidth}{!}{%
\begin{tabular}{|l|l|l|l|}
\hline
california\_wildfire                               & 2018-01-01               & 2021-12-31              & Environment                                                    \\ \hline
cameron\_outcome\_referendum                       & 2018-01-02               & 2021-12-20              & Politics, Social                                               \\ \hline
capital\_punishment                                & 2018-01-01               & 2021-12-31              & Human\_Rights, Social                                          \\ \hline
cathedral\_notre\_dame                             & 2018-01-02               & 2021-12-31              & Art\_Culture\_Sport, Environment, Social                       \\ \hline
charlie\_hebdo                                     & 2018-01-02               & 2021-12-31              & Art\_Culture\_Sport, Human\_Rights, Politics, Religion, Social \\ \hline
charlottesville\_rally\_unite                      & 2018-01-02               & 2021-12-31              & Human\_Rights, Social                                          \\ \hline
chemtrails                                         & 2018-01-01               & 2021-12-31              & Environment, Social, Tech\_Sci                                 \\ \hline
climate\_warming                                   & 2018-01-01               & 2021-12-31              & Environment, Social                                            \\ \hline
co2\_emissions                                     & 2018-01-01               & 2021-12-31              & Environment, Politics, Tech\_Sci                               \\ \hline
coach\_k                                           & 2018-01-01               & 2021-12-31              & Art\_Culture\_Sport                                            \\ \hline
colombia\_farc                                     & 2018-01-01               & 2021-12-31              & Politics                                                       \\ \hline
colorado\_shooting               & 2018-01-02 & 2021-12-31 & Social                                                            \\ \hline
confederate\_statue\_removed     & 2018-01-04 & 2021-12-31 & Art\_Culture\_Sport, Human\_Rights, Social                        \\ \hline
contest\_nobel\_prize\_winner    & 2018-01-17 & 2021-12-16 & Art\_Culture\_Sport, Tech\_Sci                                    \\ \hline
correctional\_prisons            & 2018-01-02 & 2021-12-31 & Human\_Rights                                                     \\ \hline
crypto\_currency\_exchange       & 2018-01-02 & 2021-12-31 & Economy, Labor, Tech\_Sci                                         \\ \hline
cuban\_embargo                   & 2018-01-02 & 2021-12-31 & Economy, Labor, Politics                                          \\ \hline
cultural\_heritage               & 2018-01-02 & 2021-12-31 & Art\_Culture\_Sport, Environment, Human\_Rights, Religion, Social \\ \hline
cyberbullying                    & 2018-01-01 & 2021-12-31 & Social, Tech\_Sci                                                 \\ \hline
cybersecurity                    & 2018-01-01 & 2021-12-31 & Politics, Social, Tech\_Sci                                       \\ \hline
cybersquatting                   & 2018-01-09 & 2021-12-23 & Economy, Labor, Social, Tech\_Sci                                 \\ \hline
dakota\_pipeline                 & 2018-01-01 & 2021-12-31 & Economy, Environment                                              \\ \hline
dakota\_standing\_rock           & 2018-01-01 & 2021-12-31 & Art\_Culture\_Sport, Environment, Human\_Rights                   \\ \hline
delhi\_pollution                 & 2018-01-01 & 2021-12-31 & Environment, Social, Tech\_Sci                                    \\ \hline
democracy\_threat                & 2018-01-01 & 2021-12-31 & Politics, Social                                                  \\ \hline
democrat\_min\_wage              & 2021-01-02 & 2021-12-31 & Economy, Human\_Rights, Labor, Politics                           \\ \hline
dieselgate                       & 2018-01-01 & 2021-12-31 & Economy, Environment, Labor, Tech\_Sci                            \\ \hline
diplomatic\_immunity             & 2018-01-01 & 2021-12-31 & Politics                                                          \\ \hline
divorce\_equality                & 2018-01-01 & 2021-12-31 & Economy, Politics, Social                                         \\ \hline
draft\_nfl                       & 2018-01-02 & 2021-12-31 & Art\_Culture\_Sport                                               \\ \hline
duncan\_dallas\_ebola            & 2018-04-12 & 2021-08-01 & Social, Health                                                    \\ \hline
duterte\_philippines             & 2018-01-02 & 2021-01-01 & Human\_Rights, Politics, Social                                   \\ \hline
e-cigarettes                     & 2018-01-02 & 2021-12-31 & Economy, Environment, Tech\_Sci, Health                           \\ \hline
early\_late\_voter               & 2018-01-03 & 2021-12-31 & Politics, Social                                                  \\ \hline
earthquake\_nepal                & 2019-01-02 & 2022-01-01 & Environment                                                       \\ \hline
efcc\_alleged\_fraud             & 2018-01-03 & 2021-12-30 & Economy, Labor                                                    \\ \hline
el\_chapo\_guzman                & 2018-01-02 & 2021-12-31 & Economy, Social, Health,                                          \\ \hline
elon\_musk\_tesla                & 2018-01-02 & 2021-12-31 & Economy, Environment, Labor, Tech\_Sci                            \\ \hline
endangered\_species              & 2018-01-01 & 2021-12-31 & Environment, Tech\_Sci, Health                                    \\ \hline
epa\_effort                      & 2018-01-03 & 2021-12-30 & Environment, Social, Health                                       \\ \hline
erdogan\_coup\_d\_etat\_attempt  & 2018-01-21 & 2021-12-06 & Politics, Social,                                                 \\ \hline
erdogan\_turkey                  & 2018-01-01 & 2021-12-31 & Human\_Rights, Politics                                           \\ \hline
eruption\_volcanic\_ash          & 2018-01-02 & 2021-12-31 & Environment                                                       \\ \hline
european\_commission             & 2018-01-01 & 2021-12-31 & Economy, Labor, Politics, Social                                  \\ \hline
fact-checking                    & 2018-01-01 & 2021-12-31 & Social                                                            \\ \hline
factory\_farming                 & 2018-01-01 & 2021-12-31 & Economy, Environment, Labor                                       \\ \hline
fadnavis\_maharashtra            & 2018-01-02 & 2021-12-31 & Environment, Politics, Social                                     \\ \hline
farmers\_irrigation\_scheme      & 2018-01-02 & 2021-12-31 & Economy, Environment, Labor, Tech\_Sci                            \\ \hline
fashion\_runway                  & 2018-01-01 & 2021-12-31 & Art\_Culture\_Sport                                               \\ \hline
fiscal\_cuts                     & 2018-01-01 & 2021-12-31 & Economy, Labor, Politics                                          \\ \hline
flat\_earth                      & 2018-01-01 & 2021-12-31 & Environment, Social, Tech\_Sci                                    \\ \hline
football\_galbraith              & 2019-01-07 & 2021-12-27 & Art\_Culture\_Sport                                               \\ \hline
ford\_kavanaugh                  & 2018-02-13 & 2021-12-30 & Human\_Rights, Social                                             \\ \hline
forest\_wildfire                 & 2018-01-02 & 2021-12-31 & Environment                                                       \\ \hline
garda\_dublin                    & 2018-01-03 & 2021-12-31 & Labor, Social                                                     \\ \hline
gay\_marriages\_ban              & 2018-01-05 & 2021-12-30 & Human\_Rights, Politics, Social                                   \\ \hline
gdpr                             & 2018-01-02 & 2022-01-01 & Human\_Rights, Politics, Social                                   \\ \hline
geert\_wilders\_netherlands      & 2018-01-03 & 2021-12-05 & Human\_Rights, Politics, Religion                                 \\ \hline
gender\_bathroom                 & 2018-01-01 & 2021-12-31 & Social                                                            \\ \hline
gender\_gap                      & 2018-01-01 & 2021-12-31 & Economy, Human\_Rights, Labor, Politics, Social                   \\ \hline
\end{tabular}
    }
\end{table}

\begin{table}[H]
\resizebox{1\textwidth}{!}{%
\begin{tabular}{|l|l|l|l|}
\hline
gender\_identity                 & 2018-01-02 & 2021-12-31 & Human\_Rights, Social                                             \\ \hline
george\_bush                     & 2018-01-02 & 2021-12-31 & Politics                                                          \\ \hline
germany\_nazi\_merkel            & 2018-01-02 & 2021-12-24 & Human\_Rights, Politics, Religion, Social                         \\ \hline
grace\_mugabe                    & 2018-01-02 & 2021-12-30 & Economy, Environment, Human\_Rights, Politics                     \\ \hline
greek\_bailout\_tsipras          & 2018-01-08 & 2021-07-16 & Economy, Labor, Politics, Social                                  \\ \hline
hackers\_disinformation          & 2018-01-03 & 2021-12-28 & Social, Tech\_Sci                                                 \\ \hline
haftar\_lybia                    & 2018-01-02 & 2021-12-31 & Politics                                                          \\ \hline
hajj\_pilgrimage                 & 2018-01-02 & 2021-12-31 & Religion, Social                                                  \\ \hline
halifax\_mass\_shooting          & 2018-01-25 & 2021-12-06 & Social                                                            \\ \hline
hamas                            & 2018-01-01 & 2021-12-31 & Human\_Rights, Politics, Religion                                 \\ \hline
harvey\_weinstein\_sexual\_abuse & 2018-01-02 & 2021-12-31 & Art\_Culture\_Sport, Human\_Rights                                \\ \hline
hate\_speech                     & 2018-01-01 & 2021-12-31 & Social                                                            \\ \hline
hezbollah\_lebanon               & 2018-01-02 & 2021-12-31 & Human\_Rights, Politics, Religion, Social                         \\ \hline
hijab\_ban                       & 2018-01-01 & 2021-12-31 & Human\_Rights, Religion, Social                                   \\ \hline
holocaust                        & 2018-01-01 & 2021-12-31 & Human\_Rights, Religion, Social                                   \\ \hline
homeless\_shelter                & 2018-01-01 & 2021-12-31 & Human\_Rights, Social, Health                                     \\ \hline
hong\_kong\_protest              & 2018-01-02 & 2021-12-31 & Human\_Rights, Politics, Social                                   \\ \hline
honolulu\_civil\_beat            & 2021-11-02 & 2021-12-31 & Art\_Culture\_Sport, Labor, Politics, Social                      \\ \hline
houthi\_yemen                    & 2018-01-02 & 2021-12-31 & Politics, Religion, Social                                        \\ \hline
humanitarian\_aid                & 2018-01-01 & 2021-12-31 & Human\_Rights                                                     \\ \hline
hurricane\_dorian                & 2021-01-02 & 2021-12-31 & Environment                                                       \\ \hline
hydrogen\_vehicles               & 2018-01-01 & 2021-12-31 & Economy, Environment, Labor, Tech\_Sci                            \\ \hline
illegal\_immigration             & 2018-01-01 & 2021-12-31 & Human\_Rights, Politics                                           \\ \hline
imran\_khan                      & 2018-01-02 & 2021-12-31 & Art\_Culture\_Sport, Politics, Social                             \\ \hline
india\_foreign\_investment       & 2018-01-02 & 2021-12-31 & Economy, Labor, Politics                                          \\ \hline
intensive\_animal\_farming       & 2018-01-04 & 2021-12-31 & Economy, Environment, Labor                                       \\ \hline
iran\_foreign\_minister\_zarif   & 2018-01-02 & 2021-12-13 & Politics                                                          \\ \hline
iraqi\_kurdish\_mosul            & 2018-01-03 & 2021-12-27 & Politics, Religion, Social                                        \\ \hline
ireland\_sinn\_fein              & 2018-01-01 & 2021-12-31 & Politics                                                          \\ \hline
jakarta\_flood                   & 2018-01-03 & 2021-12-27 & Environment                                                       \\ \hline
jamal\_khashoggi                 & 2018-01-01 & 2021-12-31 & Art\_Culture\_Sport, Human\_Rights, Politics                      \\ \hline
jeffrey\_epstein                 & 2018-01-01 & 2021-12-31 & Human\_Rights                                                     \\ \hline
jeremy\_corbyn\_labour           & 2018-01-02 & 2021-12-31 & Labor, Politics, Social                                           \\ \hline
john\_mccain                     & 2018-01-01 & 2021-12-31 & Politics                                                          \\ \hline
joyce\_marcel                            & 2018-01-04 & 2021-12-31 & Art\_Culture\_Sport                             \\ \hline
julian\_assange\_wikileaks               & 2018-01-02 & 2021-12-31 & Art\_Culture\_Sport, Economy, Politics, Social  \\ \hline
kabila\_congo                            & 2018-01-01 & 2021-12-31 & Politics                                        \\ \hline
karnataka\_assembly\_poll                & 2018-01-01 & 2021-12-30 & Politics                                        \\ \hline
kayapo                                   & 2018-01-03 & 2021-12-31 & Environment, Human\_Rights                      \\ \hline
kenney\_elected\_mayor\_philadelphia     & 2018-01-20 & 2021-12-31 & Politics, Social                                \\ \hline
kiev\_donetsk\_separatists               & 2018-01-13 & 2021-12-31 & Politics, Social                                \\ \hline
kiir\_machar\_south\_sudan               & 2018-01-03 & 2021-12-31 & Politics                                        \\ \hline
kilauea\_eruption                        & 2018-01-01 & 2021-12-31 & Environment, Tech\_Sci                          \\ \hline
kim\_jong\_un                            & 2018-01-01 & 2021-12-31 & Human\_Rights, Politics                         \\ \hline
klopp\_liverpool                         & 2018-01-02 & 2021-12-31 & Art\_Culture\_Sport                             \\ \hline
labor\_movement                          & 2018-01-02 & 2021-12-31 & Economy, Labor, Politics                        \\ \hline
lahore\_rape                             & 2018-01-01 & 2021-12-26 & Human\_Rights                                   \\ \hline
lee\_kuala\_lumpur                       & 2018-01-02 & 2021-12-31 & Social                                          \\ \hline
legalize\_prostitution                   & 2018-01-05 & 2021-12-30 & Social                                          \\ \hline
leo\_varadkar\_taoiseach                 & 2018-01-02 & 2021-12-31 & Human\_Rights, Politics, Social                 \\ \hline
lgbt\_discrimination                     & 2018-01-02 & 2021-12-31 & Human\_Rights, Social                           \\ \hline
london\_mayor\_re-election\_bid          & 2018-02-01 & 2021-11-15 & Politics, Social                                \\ \hline
louisiana\_parish\_arrested              & 2018-01-02 & 2021-12-31 & Social                                          \\ \hline
lung\_cancer                             & 2018-01-01 & 2021-12-31 & Social, Tech\_Sci, Health                       \\ \hline
macron\_france                           & 2018-01-01 & 2021-12-31 & Politics                                        \\ \hline
maduro\_venezuela                        & 2018-01-02 & 2021-12-29 & Politics                                        \\ \hline
marco\_rubio\_debate                     & 2021-01-03 & 2021-12-25 & Politics                                        \\ \hline
marijuana\_legalization                  & 2018-01-02 & 2022-01-01 & Economy, Politics, Social, Health               \\ \hline
marine\_corps                            & 2018-01-01 & 2021-12-31 & Labor, Social                                   \\ \hline
\end{tabular}
    }
\end{table}

\begin{table}[H]
\resizebox{1\textwidth}{!}{%
\begin{tabular}{|l|l|l|l|}
\hline
marine\_le\_pen                          & 2018-01-02 & 2021-12-31 & Politics                                        \\ \hline
mars\_mission                            & 2018-01-01 & 2021-12-31 & Environment, Tech\_Sci                          \\ \hline
mars\_spacecraft\_mission                & 2018-01-02 & 2021-12-31 & Environment, Tech\_Sci                          \\ \hline
martin\_luther\_king                     & 2018-01-02 & 2022-01-01 & Human\_Rights, Politics, Social                 \\ \hline
maryam\_nawaz                            & 2018-01-01 & 2021-12-31 & Politics                                        \\ \hline
mass\_shootings                          & 2018-01-02 & 2022-01-01 & Social                                          \\ \hline
measles                                  & 2018-01-01 & 2021-12-31 & Social, Tech\_Sci, Health                       \\ \hline
meghan\_harry                            & 2018-01-02 & 2022-01-01 & Politics, Social                                \\ \hline
merkel\_germany                          & 2018-01-01 & 2021-12-31 & Politics                                        \\ \hline
metoo                                    & 2018-01-01 & 2021-12-31 & Human\_Rights, Labor, Social                    \\ \hline
metric\_tonnes\_waste                    & 2018-01-02 & 2021-12-30 & Environment, Tech\_Sci                          \\ \hline
mexican\_migrants                        & 2018-01-01 & 2021-12-31 & Human\_Rights, Politics, Social                 \\ \hline
mexico\_wall                             & 2018-01-01 & 2021-12-31 & Human\_Rights, Politics, Social                 \\ \hline
mh370                                    & 2019-01-02 & 2021-12-29 & Social                                          \\ \hline
michael\_brown\_shooting                 & 2018-01-03 & 2021-12-31 & Human\_Rights, Social                           \\ \hline
michael\_cohen\_lawyer                   & 2018-01-05 & 2021-12-31 & Politics                                        \\ \hline
migration\_pact                          & 2018-01-05 & 2021-12-31 & Human\_Rights, Politics, Social                 \\ \hline
mike\_pence\_indiana                     & 2018-01-02 & 2021-12-20 & Politics                                        \\ \hline
mindanao\_martial\_law                   & 2018-01-01 & 2021-12-31 & Human\_Rights, Politics, Social, Health         \\ \hline
minimum\_wage                            & 2018-01-02 & 2021-12-31 & Economy, Human\_Rights, Labor, Politics, Social \\ \hline
mission\_moon                            & 2018-01-01 & 2021-12-31 & Environment, Tech\_Sci                          \\ \hline
modis\_narendra                          & 2018-01-05 & 2021-12-29 & Politics                                        \\ \hline
morsi\_sisi\_egypt                       & 2018-01-09 & 2021-12-30 & Politics, Religion                              \\ \hline
mueller\_probe                           & 2018-01-02 & 2021-12-30 & Politics                                        \\ \hline
muslim\_brotherhood                      & 2018-01-01 & 2021-12-31 & Religion                                        \\ \hline
nafta\_trade                             & 2018-01-01 & 2021-12-31 & Economy, Labor, Politics                        \\ \hline
native\_american\_indigenous             & 2018-01-01 & 2021-12-31 & Human\_Rights                                   \\ \hline
natural\_gas\_prices                     & 2018-01-01 & 2021-12-31 & Economy, Environment, Labor, Politics           \\ \hline
nauru\_refugees                          & 2018-01-02 & 2021-12-31 & Human\_Rights                                   \\ \hline
nelson\_bay\_cup                         & 2018-01-13 & 2021-12-31 & Art\_Culture\_Sport                             \\ \hline
netanyahu                                & 2018-01-01 & 2021-12-31 & Politics                                        \\ \hline
nicola\_sturgeon\_scotland\_independence & 2018-01-02 & 2021-12-31 & Politics, Social                                \\ \hline
nigel\_farage\_ukip                      & 2018-01-02 & 2021-12-28 & Politics, Social                                \\ \hline
nikolas\_cruz                            & 2018-01-09 & 2021-12-31 & Social                                          \\ \hline
nitish\_kumar\_bihar                     & 2018-01-02 & 2021-12-31 & Politics                                        \\ \hline
npa\_rebels                              & 2018-01-01 & 2021-12-31 & Politics                                        \\ \hline
nuclear\_war                             & 2018-01-01 & 2021-12-31 & Environment, Politics, Social, Tech\_Sci        \\ \hline
obrador\_mexico                          & 2018-01-01 & 2021-12-31 & Politics                                        \\ \hline
ocean\_fishing                           & 2018-01-02 & 2021-12-31 & Economy, Environment, Labor                     \\ \hline
off\_peak\_season\_travel                & 2018-01-01 & 2021-12-31 & Art\_Culture\_Sport, Economy, Environment       \\ \hline
offshore\_wind                           & 2018-01-01 & 2021-12-31 & Economy, Environment, Tech\_Sci                 \\ \hline
operation\_varsity\_blues                & 2018-06-16 & 2021-12-31 & Art\_Culture\_Sport, Labor, Social              \\ \hline
opioid\_drug\_crisis                     & 2018-01-01 & 2021-12-31 & Social, Health                                  \\ \hline
organ\_trade                             & 2018-01-01 & 2021-12-31 & Economy, Human\_Rights, Health                  \\ \hline
pacific\_solution                        & 2018-01-01 & 2021-12-31 & Human\_Rights, Politics                         \\ \hline
panama\_papers                           & 2018-01-02 & 2021-12-31 & Economy                                         \\ \hline
paul\_manafort                    & 2018-08-02 & 2021-12-30 & Politics                                           \\ \hline
pension\_retirement\_age          & 2018-01-01 & 2021-12-31 & Economy, Labor, Politics, Social, Health           \\ \hline
pkk                               & 2018-01-02 & 2021-12-31 & Human\_Rights, Politics, Social                    \\ \hline
planned\_pregnancy                & 2018-01-02 & 2022-01-01 & Human\_Rights, Social, Tech\_Sci, Health           \\ \hline
plastic\_surgery                  & 2018-01-02 & 2021-12-31 & Social, Tech\_Sci, Health                          \\ \hline
ponte\_morandi                    & 2018-01-03 & 2021-12-31 & Social                                             \\ \hline
portland\_standoff                & 2018-01-03 & 2021-12-12 & Human\_Rights, Social                              \\ \hline
protest\_sign\_mou                & 2018-01-09 & 2021-12-22 & Social                                             \\ \hline
rajapaksa\_sri\_lanka             & 2018-01-02 & 2021-12-31 & Politics                                           \\ \hline
rajasthan\_mps                    & 2018-01-07 & 2021-12-31 & Politics                                           \\ \hline
rajya\_sabha\_elections\_bjp      & 2018-01-02 & 2021-12-30 & Politics, Social                                   \\ \hline
rakhine\_rohingya\_myanmar        & 2018-01-01 & 2021-12-31 & Human\_Rights, Religion, Social                    \\ \hline
ramaphosa\_south\_africa          & 2018-01-01 & 2021-12-31 & Human\_Rights, Labor, Politics, Social             \\ \hline
randolph\_holhut                  & 2018-02-21 & 2021-09-30 & Art\_Culture\_Sport                                \\ \hline
\end{tabular}
    }
\end{table}

\begin{table}[H]
\resizebox{1\textwidth}{!}{%
\begin{tabular}{|l|l|l|l|}
\hline
ransomware                        & 2018-01-01 & 2021-12-31 & Economy, Tech\_Sci                                 \\ \hline
rauner\_illinois                  & 2018-01-01 & 2021-12-22 & Politics                                           \\ \hline
recreational\_cannabis            & 2018-01-01 & 2021-12-31 & Politics, Social, Health                           \\ \hline
religion\_freedom                 & 2018-01-01 & 2021-12-31 & Human\_Rights, Religion                            \\ \hline
reynolds\_mourned                 & 2018-07-04 & 2021-12-31 & Labor, Social                                      \\ \hline
rock\_n\_roll\_savannah\_marathon & 2018-01-04 & 2021-12-14 & Art\_Culture\_Sport, Social                        \\ \hline
roe\_v.\_wade\_case               & 2018-01-02 & 2021-12-31 & Human\_Rights, Social                              \\ \hline
ryanair\_pilot\_strike            & 2018-01-04 & 2019-10-03 & Labor                                              \\ \hline
salman\_saudi\_arabia             & 2018-01-01 & 2021-12-31 & Human\_Rights, Politics, Religion,                 \\ \hline
santos\_colombia                  & 2018-01-02 & 2021-12-31 & Politics                                           \\ \hline
sargsyan\_armenia                 & 2018-01-02 & 2021-12-31 & Politics                                           \\ \hline
scientist\_hansen\_nasa           & 2018-01-06 & 2021-12-29 & Art\_Culture\_Sport, Environment, Labor, Tech\_Sci \\ \hline
scott\_morrison\_kabul            & 2018-02-18 & 2021-12-30 & Human\_Rights, Politics                            \\ \hline
scott\_walker\_wisconsin          & 2018-01-02 & 2021-12-31 & Politics                                           \\ \hline
self-driving\_car                 & 2018-01-01 & 2021-12-31 & Economy, Labor, Social, Tech\_Sci                  \\ \hline
shiv\_sena\_maharashtra           & 2018-01-02 & 2021-12-31 & Human\_Rights, Politics                            \\ \hline
sirisena\_sri\_lanka              & 2018-01-01 & 2021-12-30 & Politics                                           \\ \hline
smart\_cities                     & 2018-01-01 & 2021-12-31 & Economy, Environment, Social, Tech\_Sci            \\ \hline
smith\_scandal\_resigns           & 2018-03-25 & 2021-11-26 & Art\_Culture\_Sport                                \\ \hline
snowden                           & 2018-01-02 & 2021-12-31 & Human\_Rights, Politics, Social, Tech\_Sci         \\ \hline
society\_civil\_servants          & 2018-01-01 & 2021-12-31 & Labor, Politics                                    \\ \hline
solar\_panels                     & 2018-01-01 & 2021-12-31 & Economy, Environment, Tech\_Sci                    \\ \hline
sonia\_rahul\_gandhi              & 2018-01-02 & 2021-12-31 & Politics                                           \\ \hline
spacex\_moon\_mission             & 2018-01-01 & 2021-12-30 & Environment, Labor, Tech\_Sci                      \\ \hline
spending\_cuts                    & 2018-01-01 & 2021-12-31 & Economy, Labor, Politics, Social                   \\ \hline
stolen\_identity                  & 2018-01-01 & 2021-12-31 & Social                                             \\ \hline
syria\_kurdish\_war               & 2018-01-02 & 2021-12-30 & Human\_Rights, Politics, Social                    \\ \hline
syrian\_refugees                  & 2018-01-01 & 2021-12-31 & Human\_Rights                                      \\ \hline
tamil\_ltte                       & 2018-01-04 & 2021-12-31 & Human\_Rights, Politics, Social                    \\ \hline
telangana\_rao\_trs               & 2018-01-02 & 2021-12-31 & Politics                                           \\ \hline
theresa\_may\_brexit              & 2018-01-02 & 2021-12-30 & Politics, Social                                   \\ \hline
tiger\_woods\_win                 & 2018-01-02 & 2021-12-31 & Art\_Culture\_Sport                                \\ \hline
tony\_abbott\_malcolm\_turnbull   & 2018-01-02 & 2022-01-01 & Politics                                           \\ \hline
tourism\_boost                    & 2018-01-01 & 2021-12-31 & Economy, Environment, Labor                        \\ \hline
tourists\_overrun                 & 2018-01-03 & 2021-12-31 & Environment, Social                                \\ \hline
tpp                               & 2018-01-02 & 2021-12-31 & Economy, Politics                                  \\ \hline
truck\_highway\_crashes           & 2018-01-02 & 2021-12-31 & Labor                                              \\ \hline
trudeau\_canada                   & 2018-01-02 & 2021-12-31 & Politics                                           \\ \hline
trump\_impeachment                & 2018-01-01 & 2021-12-31 & Politics                                           \\ \hline
tsai\_ing-wen\_taiwan             & 2018-01-01 & 2021-12-31 & Human\_Rights, Politics, Social                    \\ \hline
tshisekedi\_congo                 & 2018-01-01 & 2021-12-31 & Politics                                           \\ \hline
tupac                             & 2018-01-26 & 2021-12-09 & Art\_Culture\_Sport                                \\ \hline
uber\_ride\_sharing               & 2018-01-02 & 2021-12-31 & Economy, Labor                                     \\ \hline
uhuru\_kenyatta                   & 2018-01-01 & 2021-12-31 & Politics                                           \\ \hline
undercover\_comey\_fbi            & 2018-02-07 & 2021-11-08 & Politics                                           \\ \hline
undp\_procurement                 & 2018-01-04 & 2021-12-31 & Economy, Human\_Rights, Labor, Politics, Social    \\ \hline
unhcr\_refugees                   & 2018-01-01 & 2021-12-31 & Human\_Rights                                      \\ \hline
vatican\_abuse                    & 2018-01-03 & 2022-01-01 & Human\_Rights, Religion                            \\ \hline
white\_supremacist                & 2018-01-01 & 2021-12-31 & Human\_Rights, Social                              \\ \hline
william\_barr\_attorney\_general  & 2018-04-30 & 2021-12-31 & Politics                                           \\ \hline
william\_kate\_middleton          & 2018-01-02 & 2021-12-31 & Politics, Social                                   \\ \hline
williams\_plead\_guilty           & 2018-01-03 & 2021-12-25 & Social                                             \\ \hline
wirecard\_scandal                 & 2019-01-02 & 2022-05-05 & Economy, Labor                                     \\ \hline
women\_abortion                   & 2018-01-02 & 2022-01-01 & Human\_Rights, Social, Health                      \\ \hline
xi\_jinping                       & 2018-01-01 & 2021-12-31 & Politics                                           \\ \hline
xinhua\_silk\_road                & 2018-01-19 & 2021-12-30 & Economy, Labor                                     \\ \hline
yakubu\_dogara\_sacked            & 2018-01-15 & 2021-12-05 & Politics                                           \\ \hline
yanukovych\_crimea                & 2018-01-03 & 2021-12-30 & Politics                                           \\ \hline
ywca                              & 2018-01-02 & 2021-12-31 & Human\_Rights, Religion, Social                    \\ \hline
zayed\_uae                        & 2018-01-01 & 2021-12-31 & Economy, Politics, Social                          \\ \hline
zika\_virus                       & 2018-01-02 & 2021-12-31 & Social, Health                                     \\ \hline
zuma\_south\_africa               & 2018-01-01 & 2021-12-31 & Human\_Rights, Politics, Social                    \\ \hline
\end{tabular}
    }
\caption{List of terms employed to perform the research for each topic together with the first and last date when a post related to each topic was found.}
\label{tab:topics}
\end{table}


\subsection*{Evaluating the relationship between topic resonance and controversy}

\subsection*{Goodness of the fitting procedure}
\label{sec:fitting_procedure}
\begin{figure}[!ht]
    \centering
    \includegraphics[scale = 0.5]{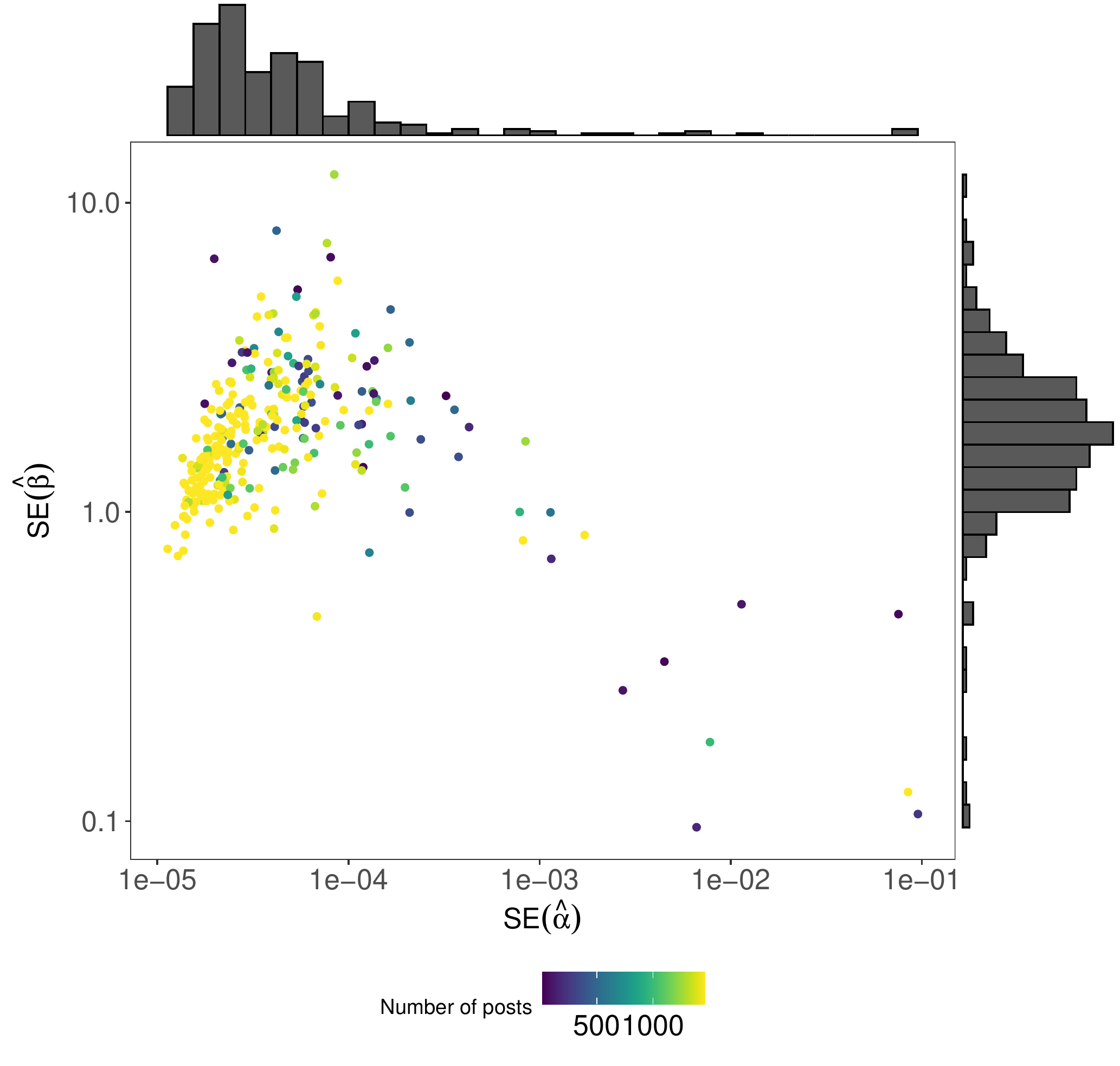}
    \caption{Joint distribution of the errors $SE(\hat{\alpha_i})$ and $SE(\hat{\beta_i})$ for each topic $i$, whose cumulative curve was estimated by means of $f_{\alpha, \beta}$. The colour of each point represent the number of posts produced by topic $i$.}
    \label{fig:error}
\end{figure}

We fit the cumulative evolution of engagement for the topics in Section \ref{sec:list_topics} with the function $f_{\alpha, \beta}$. The fitting procedure produces, for each topic $i$, the tuples $\left(\hat{\alpha_i}, \hat{\beta_i}\right)$ and $\left(SE(\hat{\alpha_i}), SE(\hat{\beta_i})\right)$ containing the estimated parameters with their standard errors, respectively. Fig. \ref{fig:error} provides a joint distribution of the errors $\left(SE(\hat{\alpha_i}), SE(\hat{\beta_i})\right)$ for each topic in relationship with the number of posts they produced. We observe how $SE(\hat{\alpha_i})$ errors follow a log-normal distribution, while $SE(\hat{\beta_i})$ errors have a normal one. We can observe a reduction in the errors for both parameters as the number of posts per topic increases. We formerly assess such relationship by computing a Spearman correlation coefficient between each standard error and the number of posts per topic, obtaining a value of $\rho(SE(\hat{\alpha_i}), posts_i) = -0.44$ and $\rho(SE(\hat{\beta_i}), posts_i) = -0.25$. We can therefore conclude that our fitting procedure provides results with a reducing error as the number of observations increases.

\subsection*{Assessing the differences of engagement behaviors across topic categories}
\label{sec:test_results}

\begin{table}[H]
\resizebox{1\textwidth}{!}{%
\begin{tabular}{|l|l|l|l|}
\hline
\textbf{Category}   & \textbf{$\alpha$}  & \textbf{$\beta$}   & \textbf{SI} \\ \hline
Art\_Culture\_Sport & 0.043 (0.1647)  & 693.8 (236)     & 0.49 (0.13)    \\ \hline
Economy             & 0.0045 (0.0022) & 752.36 (241.45) & 0.48 (0.14)    \\ \hline
Environment         & 0.0215 (0.1214) & 761.41 (207.81) & 0.47 (0.11)    \\ \hline
Human\_Rights       & 0.0244 (0.113)  & 765.48 (223.24) & 0.47 (0.14)    \\ \hline
Labor               & 0.0137 (0.0618) & 715.41 (286.41) & 0.49 (0.15)    \\ \hline
Politics            & 0.0192 (0.0953) & 711.78 (243.12) & 0.5 (0.15)     \\ \hline
Religion            & 0.0405 (0.1906) & 786.5 (184.07)  & 0.46 (0.12)    \\ \hline
Social              & 0.024 (0.1182)  & 728.58 (204.3)  & 0.49 (0.12)    \\ \hline
Tech\_Sci           & 0.004 (0.0013)  & 801.76 (187.67) & 0.46 (0.11)    \\ \hline
Health              & 0.0045 (0.0016) & 692.03 (128.7)  & 0.52 (0.08)    \\ \hline
\end{tabular}
}
\caption{Summary of $\alpha$, $\beta$ and Speed Index mean values (and SD) per topic category. }
\end{table}

\begin{table}[H]
\begin{tabular}{|l|l|l|}
\hline
\textbf{Category}   & \textbf{Rho} & \textbf{p.value} \\ \hline
All                 & -0.26        & 0                \\ \hline
Art\_Culture\_Sport & -0.2         & 0.2631           \\ \hline
Economy             & -0.2         & 0.1665           \\ \hline
Environment         & -0.21        & 0.1379           \\ \hline
Human\_Rights       & -0.29        & 0.0067           \\ \hline
Labor               & -0.35        & 0.0162           \\ \hline
Politics            & -0.36        & 0                \\ \hline
Religion            & -0.12        & 0.5308           \\ \hline
Social              & -0.23        & 0.0057           \\ \hline
Tech\_Sci           & -0.21        & 0.2256           \\ \hline
Health              & -0.45        & 0.0267           \\ \hline
\end{tabular}
\caption{Spearman's Rho between Speed and Love-Hate Score per category (CI = 0.95). For readability, 0 represents values lower than 0.0001.}
\label{tab:corr_speed_LH}
\end{table}

\begin{table}[H]
\resizebox{1\textwidth}{!}{%
\begin{tabular}{|l|l|l|l|l|l|l|l|l|l|l|}
\hline
                  & \textbf{A\_C\_S} & \textbf{Econ} & \textbf{Env} & \textbf{H\_R}   & \textbf{Labor} & \textbf{Politics} & \textbf{Religion} & \textbf{Social} & \textbf{Tech}   & \textbf{Health} \\ \hline
\textbf{A\_C\_S}  & \textbf{}        & 0.3621        & 0.9226       & 0.3671          & 0.7594         & 0.5024            & 0.8931            & 0.8646          & 0.1195          & 0.9688          \\ \hline
\textbf{Econ}     & 0.3621           & \textbf{}     & 0.2994       & 0.0104          & 0.4292         & 0.0146            & 0.3788            & 0.0793          & 0.3247          & 0.3496          \\ \hline
\textbf{Env}      & 0.9226           & 0.2994        & \textbf{}    & 0.1287          & 0.9135         & 0.2025            & 0.992             & 0.5873          & 0.0546          & 0.9688          \\ \hline
\textbf{H\_R}     & 0.3671           & 0.0104        & 0.1287       & \textbf{}       & 0.134          & 0.6761            & 0.2437            & 0.2236          & \textbf{0.0009} & 0.2014          \\ \hline
\textbf{Labor}    & 0.7594           & 0.4292        & 0.9135       & 0.134           & \textbf{}      & 0.1818            & 0.8575            & 0.4937          & 0.0874          & 0.9322          \\ \hline
\textbf{Politics} & 0.5024           & 0.0146        & 0.2025       & 0.6761          & 0.1818         & \textbf{}         & 0.335             & 0.3465          & 0.0018          & 0.2755          \\ \hline
\textbf{Religion} & 0.8931           & 0.3788        & 0.992        & 0.2437          & 0.8575         & 0.335             & \textbf{}         & 0.6836          & 0.0976          & 0.9347          \\ \hline
\textbf{Social}   & 0.8646           & 0.0793        & 0.5873       & 0.2236          & 0.4937         & 0.3465            & 0.6836            & \textbf{}       & 0.011           & 0.6151          \\ \hline
\textbf{Tech}     & 0.1195           & 0.3247        & 0.0546       & \textbf{0.0009} & 0.0874         & 0.0018            & 0.0976            & 0.011           & \textbf{}       & 0.08            \\ \hline
\textbf{Health}   & 0.9688           & 0.3496        & 0.9688       & 0.2014          & 0.9322         & 0.2755            & 0.9347            & 0.6151          & 0.08            & \textbf{}       \\ \hline
\end{tabular}
}
\caption{p-values of the two-tailed Mann–Whitney U tests performed on the average $\alpha$ parameter value between categories (CI = 0.95). Bold values represent the value for which the null hypothesis was rejected. }
\end{table}

\begin{table}[H]
\resizebox{1\textwidth}{!}{%
\begin{tabular}{|l|l|l|l|l|l|l|l|l|l|l|}
\hline
                  & \textbf{A\_C\_S} & \textbf{Econ} & \textbf{Env} & \textbf{H\_R} & \textbf{Labor} & \textbf{Politics} & \textbf{Religion} & \textbf{Social} & \textbf{Tech} & \textbf{Health} \\ \hline
\textbf{A\_C\_S}  & \textbf{}        & 0.1688        & 0.171        & 0.2124        & 0.5789         & 0.7218            & 0.2055            & 0.5985          & 0.0532        & 0.8082          \\ \hline
\textbf{Econ}     & 0.1688           & \textbf{}     & 0.952        & 0.7537        & 0.6159         & 0.1358            & 0.9875            & 0.181           & 0.5199        & 0.0567          \\ \hline
\textbf{Env}      & 0.171            & 0.952         & \textbf{}    & 0.718         & 0.6412         & 0.0932            & 0.9759            & 0.1386          & 0.4838        & 0.0315          \\ \hline
\textbf{H\_R}     & 0.2124           & 0.7537        & 0.718        & \textbf{}     & 0.6522         & 0.1274            & 0.7229            & 0.1808          & 0.2939        & 0.0599          \\ \hline
\textbf{Labor}    & 0.5789           & 0.6159        & 0.6412       & 0.6522        & \textbf{}      & 0.5506            & 0.5529            & 0.7086          & 0.2948        & 0.2793          \\ \hline
\textbf{Politics} & 0.7218           & 0.1358        & 0.0932       & 0.1274        & 0.5506         & \textbf{}         & 0.1676            & 0.7814          & 0.0274        & 0.3969          \\ \hline
\textbf{Religion} & 0.2055           & 0.9875        & 0.9759       & 0.7229        & 0.5529         & 0.1676            & \textbf{}         & 0.1816          & 0.5318        & 0.0418          \\ \hline
\textbf{Social}   & 0.5985           & 0.181         & 0.1386       & 0.1808        & 0.7086         & 0.7814            & 0.1816            & \textbf{}       & 0.0338        & 0.2407          \\ \hline
\textbf{Tech}     & 0.0532           & 0.5199        & 0.4838       & 0.2939        & 0.2948         & 0.0274            & 0.5318            & 0.0338          & \textbf{}     & 0.0076          \\ \hline
\textbf{Health}   & 0.8082           & 0.0567        & 0.0315       & 0.0599        & 0.2793         & 0.3969            & 0.0418            & 0.2407          & 0.0076        & \textbf{}       \\ \hline
\end{tabular}
}
\caption{p-values of the two-tailed Mann–Whitney U tests performed on the average $\beta$ parameter value between categories (CI = 0.95). Bold values represent the value for which the null hypothesis was rejected. }
\end{table}

\begin{table}[H]
\resizebox{1\textwidth}{!}{%
\begin{tabular}{|l|l|l|l|l|l|l|l|l|l|l|}
\hline
                  & \textbf{A\_C\_S} & \textbf{Econ} & \textbf{Env} & \textbf{H\_R} & \textbf{Labor} & \textbf{Politics} & \textbf{Religion} & \textbf{Social} & \textbf{Tech} & \textbf{Health} \\ \hline
\textbf{A\_C\_S}  & \textbf{}        & 0.4121        & 0.3312       & 0.3731        & 0.5789         & 0.869             & 0.4838            & 0.8942          & 0.1862        & 0.4756          \\ \hline
\textbf{Econ}     & 0.4121           & \textbf{}     & 0.9253       & 0.9877        & 0.9856         & 0.2284            & 0.9212            & 0.3115          & 0.5664        & 0.0496          \\ \hline
\textbf{Env}      & 0.3312           & 0.9253        & \textbf{}    & 0.8325        & 0.908          & 0.1166            & 0.9037            & 0.1748          & 0.6437        & 0.0223          \\ \hline
\textbf{H\_R}     & 0.3731           & 0.9877        & 0.8325       & \textbf{}     & 0.9216         & 0.1215            & 0.9618            & 0.2021          & 0.5007        & 0.0447          \\ \hline
\textbf{Labor}    & 0.5789           & 0.9856        & 0.908        & 0.9216        & \textbf{}      & 0.3042            & 0.8918            & 0.4106          & 0.5814        & 0.0888          \\ \hline
\textbf{Politics} & 0.869            & 0.2284        & 0.1166       & 0.1215        & 0.3042         & \textbf{}         & 0.2611            & 0.7012          & 0.0617        & 0.3407          \\ \hline
\textbf{Religion} & 0.4838           & 0.9212        & 0.9037       & 0.9618        & 0.8918         & 0.2611            & \textbf{}         & 0.3424          & 0.6332        & 0.0733          \\ \hline
\textbf{Social}   & 0.8942           & 0.3115        & 0.1748       & 0.2021        & 0.4106         & 0.7012            & 0.3424            & \textbf{}       & 0.0784        & 0.1822          \\ \hline
\textbf{Tech}     & 0.1862           & 0.5664        & 0.6437       & 0.5007        & 0.5814         & 0.0617            & 0.6332            & 0.0784          & \textbf{}     & 0.0103          \\ \hline
\textbf{Health}   & 0.4756           & 0.0496        & 0.0223       & 0.0447        & 0.0888         & 0.3407            & 0.0733            & 0.1822          & 0.0103        & \textbf{}       \\ \hline
\end{tabular}
}
\caption{p-values of the two-tailed Mann–Whitney U tests performed on the average Speed Index between categories (CI = 0.95). Bold values represent the value for which the null hypothesis was rejected.}
\end{table}

\begin{table}[H]
\resizebox{1\textwidth}{!}{%
\begin{tabular}{|l|l|l|l|l|l|l|l|l|l|l|}
\hline
                  & \textbf{A\_C\_S} & \textbf{Econ} & \textbf{Env}    & \textbf{H\_R}   & \textbf{Labor} & \textbf{Politics} & \textbf{Religion} & \textbf{Social} & \textbf{Tech}   & \textbf{Health} \\ \hline
\textbf{A\_C\_S}  & \textbf{}        & 0.0369        & 0.2472          & \textbf{0.0001} & 0.009          & \textbf{0.0001}   & 0.0048            & \textbf{0.0005} & 0.3624          & 0.0656          \\ \hline
\textbf{Econ}     & 0.0369           & \textbf{}     & 0.1372          & 0.0023          & 0.4167         & 0.0089            & 0.0792            & 0.0501          & 0.0777          & 0.9081          \\ \hline
\textbf{Env}      & 0.2472           & 0.1372        & \textbf{}       & \textbf{0}      & 0.0238         & \textbf{0}        & 0.004             & \textbf{0.0003} & 0.7374          & 0.1817          \\ \hline
\textbf{H\_R}     & \textbf{0.0001}  & 0.0023        & \textbf{0}      & \textbf{}       & 0.0401         & 0.4758            & 0.5167            & 0.1535          & \textbf{0}      & 0.0275          \\ \hline
\textbf{Labor}    & 0.009            & 0.4167        & 0.0238          & 0.0401          & \textbf{}      & 0.107             & 0.3184            & 0.3601          & 0.0121          & 0.5761          \\ \hline
\textbf{Politics} & \textbf{0.0001}  & 0.0089        & \textbf{0}      & 0.4758          & 0.107          & \textbf{}         & 0.9014            & 0.4024          & \textbf{0}      & 0.0708          \\ \hline
\textbf{Religion} & 0.0048           & 0.0792        & 0.004           & 0.5167          & 0.3184         & 0.9014            & \textbf{}         & 0.7623          & 0.0013          & 0.1672          \\ \hline
\textbf{Social}   & \textbf{0.0005}  & 0.0501        & \textbf{0.0003} & 0.1535          & 0.3601         & 0.4024            & 0.7623            & \textbf{}       & \textbf{0.0002} & 0.1763          \\ \hline
\textbf{Tech}     & 0.3624           & 0.0777        & 0.7374          & \textbf{0}      & 0.0121         & \textbf{0}        & 0.0013            & \textbf{0.0002} & \textbf{}       & 0.1064          \\ \hline
\textbf{Health}   & 0.0656           & 0.9081        & 0.1817          & 0.0275          & 0.5761         & 0.0708            & 0.1672            & 0.1763          & 0.1064          & \textbf{}       \\ \hline
\end{tabular}
}
\caption{p-values of the two-tailed Mann–Whitney U tests performed on the average Love-Hate Score between categories (CI = 0.95). Bold values represent the value for which the null hypothesis was rejected.}
\label{tab:LH_test}
\end{table}

\begin{table}[H]
\begin{tabular}{|l|l|l|l|l|l|l|}
\hline
                  & \textbf{A\_C\_S} & \textbf{Env} & \textbf{Tech} & \textbf{Politics} & \textbf{Social} & \textbf{H\_R} \\ \hline
\textbf{A\_C\_S}  &                  &              &               & \textbf{0.0001}   & \textbf{0.0003} & \textbf{0}    \\ \hline
\textbf{Env}      &                  &              &               & \textbf{0}        & \textbf{0.0002} & \textbf{0}    \\ \hline
\textbf{Tech}     &                  &              &               & \textbf{0}        & \textbf{0.0001} & \textbf{0}    \\ \hline
\textbf{Politics} & 0.9999           & 1            & 1             &                   &                 &               \\ \hline
\textbf{Social}   & 0.9998           & 0.9998       & 0.9999        &                   &                 &               \\ \hline
\textbf{H\_R}     & 1                & 1            & 1             &                   &                 &               \\ \hline
\end{tabular}
\caption{p-values of Mann–Whitney U test on LH mean value between categories for which the null hypotesis was rejected in Table \ref{tab:LH_test} (H1: $\mu_{r} > \mu_{c}$, where r and c represent row and column category; Conf. Level = 0.95). Bold values represent the value for which the null hypothesis was rejected.}
\end{table}

\end{document}